\RequirePackage{fix-cm}
\documentclass[preprint,12pt,number]{elsarticle}
\biboptions{square,sort&compress}
\usepackage{amsmath,amssymb}
\usepackage{booktabs}
\usepackage{graphicx}
\usepackage{cleveref}
\usepackage{algorithm}
\usepackage{algpseudocode}
\usepackage{xcolor}
\usepackage{lineno}
\usepackage{csquotes}
\usepackage{multirow}
\usepackage{array}
\usepackage{tabularx}
\usepackage{url}

\newcommand{\vect}[1]{\boldsymbol{#1}}

\journal{Transportation Research Part C: Emerging Technologies}

\begin{document}

\begin{frontmatter}

\title{Dynamic Active Fleet Sizing of Heterogeneous UAV Relays for Post-Disaster Communication Networks}

\author[aff1]{Jimin Choi\corref{cor1}}
\ead{jiminch@umich.edu}
% ORCID: 0009-0008-2457-6705

\author[aff2]{Max Z. Li}
\ead{maxzli@umich.edu}
% ORCID: 0000-0001-8969-6259

\cortext[cor1]{Corresponding author}

\address[aff1]{Department of Aerospace Engineering, University of Michigan, Ann Arbor, MI 48109, USA}
\address[aff2]{Department of Aerospace Engineering, Department of Civil and Environmental Engineering, and Department of Industrial and Operations Engineering, University of Michigan, Ann Arbor, MI 48109, USA}

\begin{abstract}
Uncrewed aerial vehicle (UAV) communication systems are increasingly used to provide temporary connectivity when ground communication infrastructure is damaged by disasters such as hurricanes and wildfires. However, the amount of aerial support required changes as ground networks recover and communication demand evolves. We focus on dynamically adjusting the number and allocation of actively deployed UAVs within a fixed available fleet.  We develop a model predictive control (MPC) framework for dynamic fleet sizing of heterogeneous UAV relays during post-disaster communication recovery. The framework accounts for communication demand, ground network recovery, heterogeneous UAV operation, UAV energy availability, and backhaul capacity, which limits how much relayed traffic can be transmitted to the external communication network. The resulting formulation is a convex quadratic program that supports efficient online fleet sizing and deployment decisions.  We evaluate the framework using seven historical disaster recovery case studies based on Federal Communications Commission reports, together with four backhaul scenarios and sensitivity analyses. Results show that required active UAV fleet size varies considerably throughout recovery, while backhaul capacity and fleet composition affect communication performance, highlighting the value of adapting active fleet size over time.
\end{abstract}

\begin{highlights}
\item A convex model predictive control (MPC) framework enables online dynamic sizing of active, operational uncrewed aerial vehicle (UAV) communication relay fleets.
\item The framework integrates heterogeneous UAV operation, energy and recharge constraints, and capacity limits on links connecting UAV relays to the external network.
\item Historical disaster recovery cases and forecast error analyses evaluate UAV operations under changing conditions.
\item Results show that UAV fleet requirements change throughout recovery and depend on both aerial and communication infrastructure capacity.
\end{highlights}


\end{frontmatter}

% \linenumbers

\section{Introduction}\label{sec:intro}
%% Change all
% add more realistic post-disaster communication degraded cases
% Show why uav is good for this (with examples)
% Persuade why dynamic fleet sizing is important in this situation

Reliable communication is essential during disaster response, while communication infrastructure is often damaged at times when communication demand surges. Hurricanes and wildfires have shown that these outages can be severe and prolonged. Due to Hurricane Maria, more than 95\% of cell sites in Puerto Rico were out of service at the worst point, and wireless restoration continued for several months \cite{fcc2018AtlanticHurricanes}. During the 2023 Maui wildfires, all 21 cell sites serving six heavily affected communities in West Maui were reported out of service on August 12 \cite{fcc2023HawaiiWildfires}. These events show that post-disaster communication loss is not limited to a short interruption. Ground communication capacity recovers over time, but the recovery process can vary significantly across disasters.

Uncrewed aerial vehicle (UAV) relay systems offer a flexible way to support communication while damaged ground infrastructure is being restored. Since UAVs can be rapidly deployed and repositioned as communication demands fluctuate, they are suitable for providing temporary coverage in areas with limited ground communication service \cite{zeng2016uav,naqvi2018drone,hayajneh2016drone}. This concept is already being used in disaster response. During the 2023 Maui wildfires, FirstNet deployed a Flying Cell on Wings (COW) together with other temporary communication assets to support first responders in affected areas \cite{firstnet2023Maui}. During Hurricane Beryl in 2024, a Flying COW was deployed in Matagorda County, Texas, to provide wireless service to a local hospital and surrounding communities \cite{att2024Beryl}. Similar systems were deployed after Hurricane Helene, with Flying COWs providing temporary cellular connectivity in eight affected cities while ground network restoration was underway \cite{att2024Helene}. These deployments demonstrate the practical use of aerial communication platforms as temporary support when ground communication infrastructure is damaged or unavailable. 

While aerial communication platforms provide flexible temporary support after disasters, the level of support required changes throughout the recovery period. Ground communication capacity fluctuates as infrastructure is repaired, and communication demand varies by location and time. As communication conditions change during recovery, the required number of UAVs also changes over time. From an operational perspective, this creates a dynamic fleet management problem where both the number of active UAVs and their allocation need to be updated as system conditions change. Therefore, UAV relay planning requires accounting for how the required fleet size changes throughout disaster recovery. Furthermore, dynamic fleet sizing becomes more challenging when operational constraints are considered. For example, UAV operation is limited by battery endurance and recharge requirements \cite{uavEnergy2024} and limited backhaul capacity can restrict the communication service provided by additional UAVs \cite{uavBackhaul2020}. These operational constraints and changing communication conditions lead to different UAV fleet requirements over the recovery period.

In this paper, we develop a model predictive control (MPC) framework for dynamic fleet sizing of heterogeneous UAV relays during post-disaster communication recovery. The framework incorporates key operational constraints, including backhaul capacity, heterogeneous UAV fleet operation, UAV energy availability, communication demand, and ground network recovery. A piecewise affine approximation of the nonlinear relay model keeps the MPC problem as a convex quadratic program, allowing efficient repeated optimization at each control step. We evaluate the framework using seven historical disaster recovery cases together with multiple backhaul scenarios, component ablations, and sensitivity analyses. The numerical study examines how changing recovery conditions and operational constraints affect UAV fleet requirements and communication service.

A preliminary version of this work was presented at the \textit{AIAA AVIATION 2026 Forum} \cite{choi2026optimal}. The present study extends the preliminary work through a more comprehensive operational model and broader numerical evaluation.

\section{Literature Review}
\label{sec:lit_review}
This section reviews prior work on post-disaster UAV communication, UAV fleet sizing, dynamic resource allocation, and operational constraints relevant to UAV relay systems.

\subsection{UAVs for Post-Disaster Recovery}
\label{sec:lit_UAV_comm}
UAVs have been studied for a wide range of disaster response operations, including monitoring, emergency logistics, and delivery of relief supplies. Recent work has developed a UAV route optimization framework for exploring unknown hazards and adaptive air-ground vehicle operations for environments under changing conditions \cite{2025bilevel,choi2026airground}. UAVs have also been used in humanitarian logistics to deliver relief supplies when ground transportation is disrupted or uncertain. Optimization and learning-based approaches have considered uncertain demand locations, damaged transportation networks, and changing travel conditions in disaster response operations \cite{ghelichi2022drone,vanSteenbergen2023relief}. These studies demonstrate the flexibility of UAVs in disaster environments where access, demand, and operating conditions can change over time.

Post-disaster communication recovery is another important application of UAV systems. Damage to cellular infrastructure, power systems, and communication links can reduce or completely interrupt ground communication service after a disaster. Existing studies have examined several technologies and network architectures for restoring connectivity, including ground, aerial, and satellite communication systems \cite{8910632,9832657}. Within these systems, UAVs can operate as temporary aerial base stations or communication relays since they can be deployed rapidly and repositioned according to communication needs \cite{naqvi2018drone}. UAV trajectory planning, scheduling, communication with surviving ground base stations, direct communication between ground devices, and relay connections through multiple UAVs have also been studied in disaster communication settings \cite{8641424}.

A large part of the UAV communication literature has focused on deployment and network design after communication infrastructure is disrupted. Small cellular networks with UAVs have been studied to restore wireless coverage in disaster areas \cite{hayajneh2016drone}. Cooperative access networks have combined UAV relays with other communication links to support emergency connectivity \cite{9013813}, and rapid deployment methods have been developed for positioning multiple UAV base stations under different disaster conditions \cite{app131910723}. Multi-tier architectures have also been proposed to combine UAVs with different communication technologies and improve connectivity in disaster recovery scenarios \cite{8904140}. These studies show that UAV communication performance depends on deployment location, network structure, and the available supporting communication infrastructure.

\subsection{UAV Fleet Sizing}
\label{sec:lit_UAV_fleet}

UAV fleet sizing has been studied in logistics, delivery, and persistent service applications, where the number of available UAVs directly affects service capacity, operating cost, and resource utilization. Early fleet sizing studies considered battery endurance and energy autonomy when determining the number of UAVs required to maintain delivery service \cite{su10093344}. Other work has examined fleet size under stochastic delivery demand, where uncertain numbers and weights of delivery requests affect the number of UAVs needed to provide reliable service \cite{HADAS2024100127}. Shared drone systems have also been studied by jointly considering fleet size and service decisions, showing the close relationship between fleet capacity, customer demand, and operating cost \cite{PEI2021102482}.

Fleet sizing has also been combined with broader deployment and operational planning decisions. Stochastic fleet deployment models have jointly determined the number and type of drones together with service schedules under uncertain demand \cite{liu2019stochasticFleet}. Related studies on persistent UAV operations have considered the number of spare UAVs required to maintain continuous service when active vehicles need to leave their tasks for battery replacement or recharging \cite{hartuv2018spareDrones}. These formulations show that UAV fleet requirements depend not only on demand but also on energy limitations, service duration, and the way available vehicles are scheduled and deployed.

\subsection{Dynamic Fleet Sizing and Resource Allocation}
\label{sec:lit_dynamic_fleet}

Although most UAV fleet sizing studies determine the fleet size in advance and keep it fixed during operation, dynamic fleet sizing can be viewed as an operations management problem where resource capacity and allocation are adjusted as demand and system conditions evolve. In these applications, the required resource level can change over time as demand and system conditions evolve. Staffing and scheduling models under nonstationary demand have examined how service capacity can be adjusted across time periods to maintain service performance while controlling resource use \cite{DEFRAEYE20164}. Optimal control has also been used to determine time-varying resource levels in queueing systems, where control decisions respond to changes in demand and backlog \cite{SEIDL2016246}. More general stochastic control formulations have extended this idea to multiple resource types under uncertain or variable demand \cite{gao2013dynamicResource}.

Similar fleet sizing and allocation problems arise in transportation and shared mobility systems. Dynamic vehicle allocation models have considered uncertain transportation requests and repeatedly updated vehicle dispatch decisions while accounting for expected future demand \cite{powell1987dynamicAllocation}. Fleet sizing has also been studied in ridesharing and bicycle sharing systems, where fleet capacity is selected to balance service availability with efficient resource use \cite{auadPerez2022,lu2016bike}. Dynamic fleet management has further considered uncertain demand and vehicle availability, with vehicle assignments and repositioning decisions updated as system conditions change \cite{hajibabai2016snowPlow}. Other work has extended these ideas to fleet sizing by using a stochastic MPC framework to determine the size and allocation of heterogeneous rail freight fleets under uncertain demand and travel times \cite{milenkovic2015mpcFleet}. 

\subsection{Operational Constraints in UAV Communication Systems}
\label{sec:lit_UAV_constraints}

UAV communication systems are subject to operational constraints that affect how much communication service can be provided over time. These constraints include differences in UAV mobility and endurance, limited onboard energy, recharge requirements, and backhaul capacity, which limits how much relayed traffic can be transmitted from the UAVs to the external communication network. Heterogeneous UAV fleets can combine platforms with different mobility, endurance, and deployment characteristics. For instance, tethered UAVs can remain active for longer periods using continuous ground power, while untethered UAVs provide greater deployment flexibility \cite{tetheredUAVReview2025}. Recent post-disaster communication studies have also considered networks that combine tethered and untethered UAVs to take advantage of these complementary operating characteristics \cite{deng2026heterogeneousUAV}.

Energy availability is an important constraint for UAV operation. Limited battery capacity affects flight duration, communication service time, and the frequency of recharge cycles. Energy-aware communication models have considered the relationship between UAV operation, communication performance, and available battery energy \cite{uavEnergy2024}. Recharge periods reduce the number of UAVs available for service, making sustained operation dependent on flight and recharge cycles.

Backhaul capacity is another important constraint in UAV communication networks. UAV placement and communication performance depend on the capacity and structure of the links connecting aerial platforms to the wider network \cite{liXu2018Backhaul}. Limited backhaul can reduce the additional communication service obtained from deploying more UAVs \cite{uavBackhaul2020}. Multi-tier communication architectures similarly show that UAV service depends on the supporting communication network as well as the deployed aerial platforms \cite{8904140}.

\section{Research Gaps and Contribution}
\label{sec:contribution}
Existing research has examined UAV communication, UAV fleet sizing, and dynamic resource allocation from different perspectives. Post-disaster communication studies have largely focused on coverage, placement, trajectory, scheduling, and network design, while UAV fleet sizing studies generally determine the required fleet in advance based on demand and operational conditions. Dynamic fleet sizing and resource allocation have been studied in broader service and transportation systems, but have received less attention in UAV applications. Moreover, operational factors such as heterogeneous UAV capabilities, energy and recharge limitations, and backhaul capacity are often considered separately. These gaps motivate a framework that dynamically adjusts UAV fleet requirements throughout disaster recovery while accounting for changing communication demand, ground network recovery, and key operational constraints. The system state evolves over time through the communication backlog and available Mobile UAV energy, while UAV deployment and communication service are adjusted at each control step. This sequential decision structure motivates the use of an MPC framework.

The main contributions of this study are as follows:
\begin{itemize}
\item We formulate dynamic UAV fleet sizing for post-disaster communication as an MPC problem. The resulting convex quadratic program can be solved efficiently at each control step, enabling repeated fleet sizing and deployment decisions as recovery conditions change.

\item We incorporate key operational constraints into the fleet sizing framework, including heterogeneous UAV operation, UAV energy and recharge dynamics, and backhaul capacity limitations.

\item We conduct extensive numerical experiments using seven historical disaster recovery cases and multiple backhaul scenarios, together with component ablations and sensitivity analyses. The results show how recovery conditions and operational constraints affect UAV fleet requirements and communication service.

\end{itemize}

\section{Methodology}
\label{sec:methodology}
This section presents the post-disaster communication system, the mathematical formulation for dynamic active fleet sizing, and the MPC solution approach.

\subsection{System Model}
\label{sec:sys_model}
The system model describes the recovering ground communication network, heterogeneous UAV relay fleet, and backhaul infrastructure considered in this study.

\subsubsection{Post-Disaster Communication System}
\label{sec:sys_problem_statement}

\begin{figure}
\centering
\includegraphics[width=\linewidth]{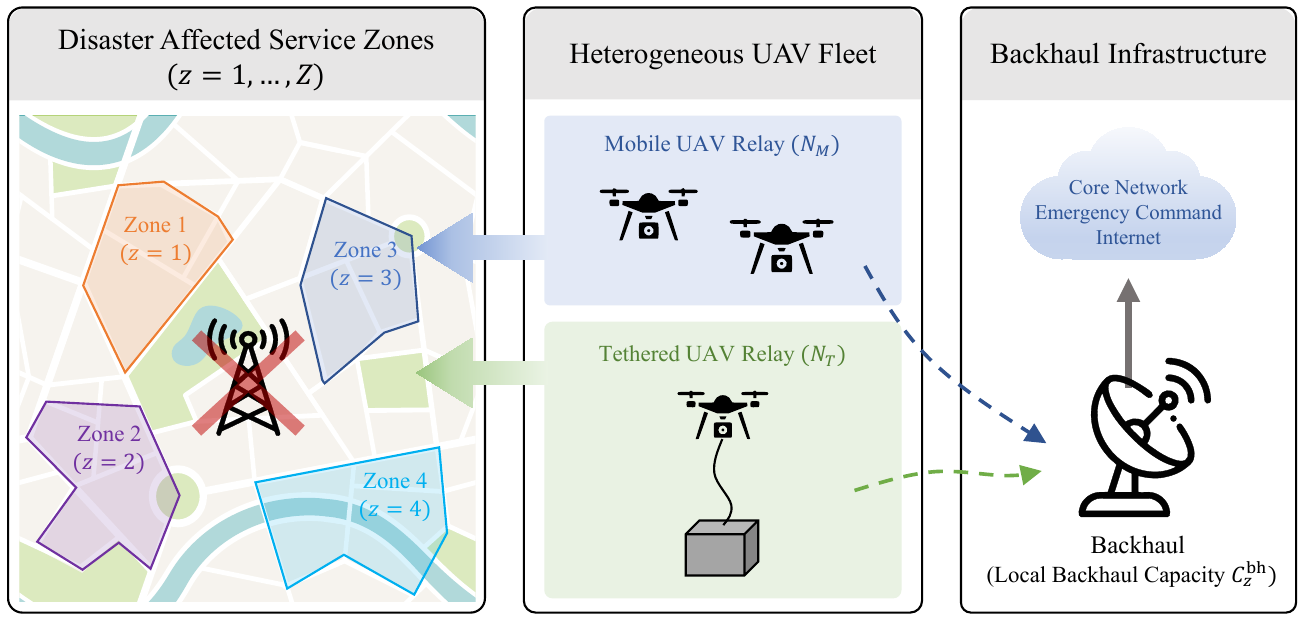}
\caption{Overview of the post-disaster communication system, including service zones, heterogeneous UAV relays, and local backhaul infrastructure.}\label{fig:system_overview}
\end{figure}

We consider a post-disaster communication system in which the affected area is divided into $Z$ service zones, indexed by $z\in\mathcal{Z}=\{1,\ldots,Z\}$, with time-varying communication demand. When a disaster happens, ground communication infrastructure may experience substantial capacity loss and subsequently recover over time. Communication demand that cannot be supported by the available ground network is supplemented by an aerial relay system composed of Mobile UAVs and Tethered UAVs. Traffic relayed by the UAVs is transmitted through a backhaul link in each service zone. As communication demand, ground network recovery, and resource availability change over time, the controller dynamically deploys UAVs across the service zones. \Cref{fig:system_overview} illustrates the overall system architecture considered in this study.

The communication load requiring aerial support is the difference between local communication demand and the service capacity available from the recovering ground network. Ground network capacity changes over time as recovery progresses, while communication demand varies across service zones and over time. Differences in population, local communication needs, and temporary demand surges associated with disaster response activities lead to different aerial support requirements across zones. As ground communication capacity recovers and local demand changes, the required UAV deployment changes accordingly. Therefore, time-varying ground recovery and varying communication demand require different UAV deployment patterns over the disaster recovery period.

The communication service provided by the aerial system depends on both UAV availability and backhaul capacity. For example, when communication demand increases in a service zone, additional Mobile UAVs are deployed to provide temporary support. However, the availability of Mobile UAVs depends on their charging cycle, while Tethered UAVs can remain active for longer periods at fixed locations and provide continuous support to zones with sustained communication needs. Additionally, the amount of traffic that can be delivered is limited by the available backhaul capacity in each zone. The controller updates UAV deployment as communication demand and resource availability change over time. The following subsections explain the ground recovery and communication demand model, the UAV relay fleet, and the backhaul infrastructure in more detail.

\subsubsection{Ground Network Recovery and Demand}
\label{sec:sys_recovery_demand}

Let $C_z^0$ denote the nominal ground communication capacity of service zone $z \in \mathcal{Z}$. A disaster may significantly reduce the available ground communication capacity, which gradually recovers as damaged communication infrastructure is restored. Post-disaster infrastructure restoration can exhibit nonlinear recovery trajectories, with recovery progressing slowly at first and accelerating as restoration activities and resources become organized \cite{zorn2015restoration,cimellaro2010resilience}. We use a logistic function to represent this recovery trajectory 
\begin{equation}
R(t)
=
\frac{K}
{1+\exp\left[-r(t-\tau)\right]},
\label{eq:recovery_logistic}
\end{equation}
where $K$ denotes the maximum recovery level, $r$ is the recovery rate, and $\tau$ is the time at which the recovery rate is highest. The corresponding ground communication capacity available in zone $z$ at time step $k$ is
\begin{equation}
G_{z,k}
=
C_z^0 R(t_k),
\label{eq:ground_capacity}
\end{equation}
where $t_k=k\Delta t$ represents the time since the start of the evaluation. The recovery parameters are fitted separately for each historical disaster case, as described in \Cref{sec:exp_historical}.

Let $D_{z,k}$ denote the communication demand in zone $z$ at time step $k$. We assume that the aerial system supports communication demand that exceeds the available ground network capacity. The remaining communication demand served by the UAV relays is defined as
\begin{equation}
\lambda_{z,k}
=
k_{\mathrm{eq}}
\max\left(D_{z,k}-G_{z,k},0\right),
\label{eq:residual_demand}
\end{equation}
where $k_{\mathrm{eq}}$ converts the unmet communication demand into the normalized service load used by the optimization model. In other words, it determines how much aerial service is required for a given amount of communication demand that cannot be supported by the ground network.

As shown in \Cref{eq:residual_demand}, we use aerial support when communication demand exceeds the available ground capacity. As the ground network recovers, the gap between communication demand and ground capacity decreases, reducing the number of UAVs required. If the ground network can fully support the local communication demand, aerial service is no longer needed.

\subsubsection{Heterogeneous UAV Relay Fleet}
\label{sec:sys_hetero_fleet}

The aerial system consists of a heterogeneous fleet of \textit{Mobile UAVs} and \textit{Tethered UAVs}. Let $N_M$ and $N_T$ denote the maximum available numbers of Mobile UAVs and Tethered UAVs, respectively. These quantities define the fleet availability limits, while the number of UAVs actively deployed can vary over time. Both UAV types provide supplemental communication service to the service zones, while they differ substantially in mobility, persistence, and energy availability.

Mobile UAVs are dynamically assigned across service zones as communication demand and ground recovery conditions change. However, they have limited battery energy and need time to recharge between deployments. The corresponding flight and recharge durations are $T_M^{\mathrm{flight}}$ and $T_M^{\mathrm{chg}}$, respectively. Moreover, Mobile UAV deployment is subject to a ramp limit $R_M$, which limits how much the deployment can increase at each control step. The ramp limit captures practical constraints on dispatching, launching, and repositioning additional Mobile UAVs. The mobility and energy constraints of Mobile UAVs allow flexible responses to changes in communication demand while linking current deployment decisions to future UAV availability. 

By comparison, Tethered UAVs are aerial platforms connected to a ground station by a physical tether. The tether provides continuous power, allowing the UAVs to remain active for long periods without returning for recharging. We assume that each Tethered UAV operates from a fixed site within a single service zone and does not move between zones. This allows Tethered UAVs to provide continuous communication support in selected service zones, but limits their ability to respond spatially when demand shifts across zones.

The relay gain parameters $g_M$ and $g_T$ represent how much communication service each Mobile UAV and Tethered UAV can provide. They map the number of active UAVs to the relay capability available to the communication system. Therefore, the aerial service depends on the number and type of UAVs deployed in each zone and the available energy of the Mobile UAV fleet. We introduce the detailed fleet states, deployment decisions, energy dynamics, and operational constraints in \Cref{sec:math_states} and \Cref{sec:math_dynamics}.

\subsubsection{Backhaul Infrastructure}
\label{sec:sys_backhaul}

Aerially relayed traffic is transmitted from the disaster area to an external communication network through a backhaul connection. Each service zone has one backhaul endpoint that supports the UAVs operating within the zone. This assumption reflects emergency communication systems where temporary communication equipment can use its own backhaul connection, such as a satellite link, when ground infrastructure is unavailable \cite{firstnetDeployables,9013813}. Multiple UAVs serving the same zone share the capacity of the local backhaul endpoint, while each zone has an independent backhaul connection.

Let $Q_{z}^{\mathrm{bh}}$ denote the effective backhaul capacity available to service zone $z$ in Mbps. We convert the backhaul capacity into the service rate used in the optimization model as
\begin{equation}
C_{z}^{\mathrm{bh}}
=
k_{\mathrm{eq}} Q_{z}^{\mathrm{bh}},
\label{eq:backhaul_normalization}
\end{equation}
where $k_{\mathrm{eq}}$ is the service scaling factor for the communication service model. The resulting $C_{z}^{\mathrm{bh}}$ represents the maximum service rate supported by the backhaul endpoint in zone $z$.

In the numerical experiments, we assume that all service zones use the same type of backhaul and the same endpoint capacity within each scenario.
\begin{equation}
Q_{z}^{\mathrm{bh}} = Q_{b}^{\mathrm{bh}},
\qquad
C_{z}^{\mathrm{bh}} = C_{b}^{\mathrm{bh}},
\qquad
\forall z,
\end{equation}
where $b$ is the selected backhaul scenario. The backhaul capacity acts as a local limit in each service zone. Once the backhaul endpoint reaches its capacity, adding more UAVs cannot increase the delivered service. We assume that each zone connects independently to the external communication network through its own backhaul link, with no additional shared capacity limit across zones.

% Other UAV communication systems may instead share a common gateway or backhaul network across multiple zones \cite{liXu2018Backhaul}. Such centralized backhaul topologies would introduce an additional aggregate or network-flow constraint and are outside the scope of the present fleet-sizing formulation.

% The numerical experiments consider multiple per-zone backhaul capacities to evaluate how local backhaul availability interacts with dynamic UAV fleet sizing. Their case-specific capacities are introduced in \Cref{sec:exp_backhaul}, while the corresponding service constraint is defined in \Cref{sec:math_dynamics}.

\subsection{Mathematical Formulation}
\label{sec:math_formulation}
The mathematical formulation defines the system states and decision variables, their dynamics and operational constraints, and the resulting MPC optimization problem.

\subsubsection{States and Decision Variables}
\label{sec:math_states}

The system is represented over discrete control intervals of duration $\Delta t$. At each control update, the MPC optimizes UAV deployment and communication service over a prediction horizon of $H$ steps. Let $h=0,\ldots,H-1$ denotes the prediction step within the current MPC problem.

We use two state variables to track the system from one control step to the next. The backlog state $x_{z,h}\ge0$ indicates communication demand that has accumulated in service zone $z$ and has not yet been served at prediction step $h$. The Mobile UAV energy state $E_h$ represents the total battery energy available across the fleet, where one unit corresponds to the energy of one fully charged Mobile UAV. The energy state is bounded by the total Mobile UAV fleet size $N_M$. In contrast, Tethered UAVs do not require an energy state as they receive continuous ground power.

The primary deployment decision for the Mobile UAV fleet is
\begin{equation}
u^M_{z,h}\ge0,
\label{eq:mobile_decision}
\end{equation}
where $u^M_{z,h}$ denotes the number of Mobile UAV equivalents assigned to service zone $z$ during prediction step $h$. The variable $u^M_{z,h}$ is modeled continuously and represents the average number of active Mobile UAVs in zone $z$ during the control interval.

Tethered UAV operation is represented by
\begin{equation}
a^T_{z,h}\ge0,
\label{eq:tethered_decision}
\end{equation}
where $a^T_{z,h}$ represents how much the Tethered UAV in zone $z$ is used during prediction step $h$. The variable $a^T_{z,h}\in[0,1]$ represents the fraction of control interval $h$ during which the Tethered UAV is active in zone $z$. Its corresponding active time is $a^T_{z,h}\Delta t$.

Let
\begin{equation}
y_{z,h}\ge0
\label{eq:service_decision}
\end{equation}
indicates the communication service delivered to zone $z$ during prediction step $h$. The delivered service depends on three factors: UAV relay capacity, backhaul capacity, and the amount of communication demand waiting to be served.

We define the effective aerial relay input in each service zone as
\begin{equation}
v_{z,h}
=
g_M u^M_{z,h}
+
g_T a^T_{z,h},
\label{eq:effective_relay_input}
\end{equation}
where $g_M$ and $g_T$ denote the service gains of Mobile UAVs and Tethered UAVs. The total active aerial deployment is similarly defined as
\begin{equation}
n_{z,h}
=
u^M_{z,h}
+
a^T_{z,h}.
\label{eq:total_aerial_deployment}
\end{equation}
In this study, dynamic fleet sizing refers to determining the active UAV fleet size at each control interval within the maximum available fleet limits. The total active fleet size is $\sum_{z=1}^{Z} n_{z,h}$, expressed in UAV equivalents averaged over the control interval. The decisions $u^M_{z,h}$ and $a^T_{z,h}$ jointly determine this fleet size and its allocation across service zones. The Mobile UAV energy state changes as active UAVs consume energy and inactive UAVs recharge on the ground. Therefore, charging is represented directly through the feasible transition of $E_h$, without a separate charging decision variable. 

% The following subsection specifies the backlog, energy, fleet, service, and backhaul constraints that govern the evolution of these states and decisions.

\subsubsection{System Dynamics and Constraints}
\label{sec:math_dynamics}

The system dynamics couple aerial relay deployment, communication service, backlog evolution, and Mobile UAV energy availability over the prediction horizon. The effective aerial relay input defined in \Cref{eq:effective_relay_input} is converted into available communication service using a saturating function
\begin{equation}
\bar{y}^{\mathrm{air}}_{z,h}
=
q_{\mathrm{ref}}
\left(
\frac{v_{z,h}}
{1+\alpha v_{z,h}}
\right),
\label{eq:aerial_service_capacity}
\end{equation}
where $\alpha$ controls the degree of service saturation. A single saturation function is applied to both UAV types when they operate in the same service zone. The nonlinear function in \Cref{eq:aerial_service_capacity} represents the actual relationship between UAV deployment and communication service. For the MPC optimization, this function is approximated using the piecewise affine model described in \Cref{sec:sol_approx}.

The service delivered in each zone cannot exceed the UAV relay capacity, the local backhaul capacity, or the amount of demand available to serve. Accordingly,
\begin{align}
0 \le y_{z,h}
&\le
\bar{y}^{\mathrm{air}}_{z,h},
\qquad
\forall z,h,
\label{eq:service_aerial_constraint}
\\
y_{z,h}
&\le
C_{z}^{\mathrm{bh}},
\qquad
\forall z,h,
\label{eq:service_backhaul_constraint}
\\
y_{z,h}
&\le
\hat{\lambda}_{z,h}
+
\frac{x_{z,h}}{\Delta t},
\qquad
\forall z,h.
\label{eq:service_demand_constraint}
\end{align}

The backhaul constraint in \eqref{eq:service_backhaul_constraint} is applied independently to each service zone. Aerial relays operating in the same zone share the capacity of the local backhaul endpoint, while unused backhaul capacity cannot be transferred between zones. This formulation is consistent with the distributed backhaul architecture described in \Cref{sec:sys_backhaul}.

The backlog in each service zone evolves according to the difference between residual communication load and delivered service. At each control step, the backlog is updated as
\begin{equation}
x_{z,k+1}
=
\max
\left(
0,
x_{z,k}
+
\Delta t
\left(
\lambda_{z,k}-y_{z,k}
\right)
\right).
\label{eq:backlog_dynamics}
\end{equation}
Within the MPC horizon, \Cref{eq:service_demand_constraint} ensures that the predicted backlog remains nonnegative. The predicted backlog is updated as
\begin{equation}
x_{z,h+1}
=
x_{z,h}
+
\Delta t
\left(
\hat{\lambda}_{z,h}-y_{z,h}
\right).
\label{eq:predicted_backlog_dynamics}
\end{equation}

Tethered UAV operation is restricted to fixed deployment sites. We assume that each Tethered UAV site serves only its assigned service zone, without explicitly modeling tether length or overlapping coverage across service zones. Let $s_z^T\in\{0,1\}$ indicate whether zone $z$ has a Tethered UAV site. The corresponding availability constraint is
\begin{equation}
0
\le
a^T_{z,h}
\le
s_z^T,
\qquad
\sum_{z=1}^{Z}s_z^T=N_T.
\label{eq:tethered_availability}
\end{equation}
The variable $a^T_{z,h}$ represents the fraction of control interval $h$ during which the Tethered UAV in zone $z$ is active, corresponding to an active time of $a^T_{z,h}\Delta t$.

The total Mobile UAV deployment at each prediction step cannot exceed the available Mobile UAV fleet. Accordingly,
\begin{equation}
\sum_{z=1}^{Z}
u^M_{z,h}
\le
N_M,
\qquad
\forall h.
\label{eq:mobile_fleet_capacity}
\end{equation}

Consistent with UAV scheduling models that represent energy limitations through flight duration and replenishment requirements, we use a simplified energy model based on Mobile UAV flight and recharge times \cite{arribas2023resupply,song2014persistent}. The energy consumption of one active Mobile UAV during a control step is defined as
\begin{equation}
\rho_M
=
\frac{\Delta t}{T_M^{\mathrm{flight}}},
\label{eq:mobile_energy_rate}
\end{equation}
and the recharge rate as
\begin{equation}
\rho_C
=
\min
\left(
\frac{\Delta t}{T_M^{\mathrm{chg}}},
1
\right).
\label{eq:mobile_recharge_rate}
\end{equation}
The energy consumed during prediction step $h$ is
\begin{equation}
d_h
=
\rho_M
\sum_{z=1}^{Z}
u^M_{z,h},
\label{eq:mobile_energy_drain}
\end{equation}
subject to the energy availability constraint
\begin{equation}
d_h
\le
E_h.
\label{eq:energy_causality}
\end{equation}
Energy consumption during a control step is limited by the energy available at the beginning of the step.

The number of Mobile UAV equivalents available for charging is
\begin{equation}
n_h^G
=
N_M
-
\sum_{z=1}^{Z}
u^M_{z,h}.
\label{eq:grounded_mobile_uavs}
\end{equation}
Let
\begin{equation}
E_h^{-}
=
E_h-d_h
\label{eq:postflight_energy}
\end{equation}
denote the energy remaining after flight. The reachable energy state at the next prediction step satisfies
\begin{equation}
E_{h+1}
\ge
E_h^{-},
\label{eq:energy_lower_bound}
\end{equation}
\begin{equation}
E_{h+1}
\le
E_h^{-}
+
\rho_C n_h^G,
\label{eq:energy_charging_bound}
\end{equation}
and
\begin{equation}
0
\le
E_{h+1}
\le
N_M.
\label{eq:energy_capacity_bound}
\end{equation}
These constraints represent fleet charging using the total energy state, without tracking the battery level or charger use of each Mobile UAV individually.

Finally, increases in Mobile UAV deployment within each service zone are limited by
\begin{equation}
u^M_{z,h}
-
u^M_{z,h-1}
\le
R_M\Delta t.
\label{eq:mobile_ramp_constraint}
\end{equation}
The constraint limits increases in Mobile UAV deployment from one control step to the next, while decreases in deployment are unrestricted. Tethered UAVs do not need the same constraint as they remain at fixed sites.

\subsubsection{MPC Formulation}
\label{sec:math_MPC}

At each control update, the MPC determines the Mobile UAV allocation, Tethered UAV activation, and delivered communication service over a prediction horizon of $H$ control steps. The optimization uses the current backlog and Mobile UAV energy state together with the predicted residual communication load $\hat{\lambda}_{z,h}$ over the horizon.

Let
\begin{equation}
\vect{x}_h
=
\begin{bmatrix}
x_{1,h} & \cdots & x_{Z,h}
\end{bmatrix}^{\mathsf T}
\end{equation}
indicate the vector of communication backlog states, and let
\begin{equation}
\vect{n}_h
=
\begin{bmatrix}
n_{1,h} & \cdots & n_{Z,h}
\end{bmatrix}^{\mathsf T}
\end{equation}
denote the total active aerial deployment across the service zones. The vector $\vect{n}_{-1}$ is the UAV deployment used in the previous control step.

The resulting optimization problem is
\begin{equation}
\begin{aligned}
\min_{\substack{
u^M_{z,h},\, a^T_{z,h},\, y_{z,h},\\
x_{z,h+1},\, E_{h+1}
}}
\quad
J
={}&
\sum_{h=0}^{H-1}
\Delta t
\Bigl[
c_x \left\|\vect{x}_{h+1}\right\|_2^2
+
c_u \left\|\vect{n}_h\right\|_2^2
+
c_s
\left\|
\vect{n}_h-\vect{n}_{h-1}
\right\|_2^2
\Bigr]
\\
&+
c_T
\left\|
\vect{x}_H
\right\|_2^2
-
\epsilon_E
\sum_{h=0}^{H-1}
E_{h+1}
\\[0.5em]
\text{subject to}\quad
& \eqref{eq:service_aerial_constraint}
  \text{--}
  \eqref{eq:service_demand_constraint},
\\
& \eqref{eq:predicted_backlog_dynamics}
  \text{--}
  \eqref{eq:mobile_fleet_capacity},
\\
& \eqref{eq:energy_causality},
  \eqref{eq:energy_lower_bound}
  \text{--}
  \eqref{eq:energy_capacity_bound},
\\
& \eqref{eq:mobile_ramp_constraint}.
\end{aligned}
\label{eq:mpc_problem}
\end{equation}

The objective function balances communication performance and aerial resource use over the prediction horizon. The backlog penalty weighted by $c_x$ promotes timely service of residual communication demand. The deployment penalty weighted by $c_u$ encourages the controller to meet communication demand with fewer active UAVs, directly influencing the fleet size selected by the MPC. The deployment variation penalty weighted by $c_s$ discourages large changes in the active UAV fleet between control steps. The terminal backlog penalty weighted by $c_T$ limits the amount of service postponed beyond the prediction horizon. The final term slightly favors solutions with more remaining Mobile UAV energy, while the small weight $\epsilon_E$ keeps this effect secondary to the main objectives. The numerical values of the objective weights are reported in \Cref{tab:case_study_parameters}.

The constraints in \Cref{eq:service_aerial_constraint,eq:service_backhaul_constraint,eq:service_demand_constraint} define the feasible communication service in each zone. Delivered service is limited by the available aerial relay capability, the capacity of the local backhaul endpoint, and the amount of residual communication traffic available for service. 
The constraints in \Cref{eq:predicted_backlog_dynamics,eq:tethered_availability,eq:mobile_fleet_capacity} describe communication backlog evolution and aerial fleet availability. The predicted backlog evolves according to the difference between residual communication load and delivered service. Tethered UAVs operate only in zones with fixed deployment sites, while the total Mobile UAV deployment cannot exceed the available fleet size.
The constraints in \Cref{eq:energy_causality,eq:energy_lower_bound,eq:energy_charging_bound,eq:energy_capacity_bound} represent the energy feasibility of the Mobile UAV fleet. The energy constraints limit Mobile UAV deployment to the energy available at the beginning of each control interval and account for energy used during flight. They also represent recharging on the ground and keep the fleet energy state within its allowed range.
Finally, \Cref{eq:mobile_ramp_constraint} limits the rate at which additional Mobile UAVs can be activated in each service zone between consecutive control intervals. 
The deployed fleet size is determined dynamically within the available fleet limits. At each prediction step, the MPC selects Mobile UAV deployment and Tethered UAV activation through $u^M_{z,h}$ and $a^T_{z,h}$. These decisions adjust the active fleet size and its allocation across service zones as communication demand, ground recovery, backhaul capacity, and Mobile UAV energy availability change over time.

\subsection{Solution Approach}
\label{sec:sol_approach}
The solution approach describes the receding horizon implementation of the MPC and the piecewise affine approximation used to preserve convexity.

\subsubsection{Receding Horizon Control}
\label{sec:sol_receding}

The MPC formulation is implemented using a receding horizon approach. At each control update, we construct the communication load expected over the prediction horizon and solve the finite horizon optimization problem in \Cref{eq:mpc_problem}. Although the MPC optimizes decisions over the full prediction horizon, only the first deployment decision is applied to the system. After this decision is implemented, the system state is updated using the actual communication service and Mobile UAV energy, and the optimization is solved again at the next control step.

Let $k$ denote the current control step. At the beginning of each control interval, the controller observes the current backlog state $\vect{x}_k$, the Mobile UAV energy state $E_k$, and the UAV deployment from the previous control step. Next, the controller constructs a forecast over $H$ steps of the residual communication load
\begin{equation}
\hat{\lambda}_{z,k+h},
\qquad
h=0,\ldots,H-1.
\label{eq:mpc_residual_forecast}
\end{equation}

The MPC uses the forecast together with the current system state and available backhaul capacity to solve the optimization problem. Solving \Cref{eq:mpc_problem} determines the UAV deployment and communication service for each step of the prediction horizon. The resulting optimal trajectories are
\begin{equation}
\left\{
u^{M*}_{z,h},
a^{T*}_{z,h},
y^*_{z,h}
\right\}_{h=0}^{H-1},
\label{eq:mpc_optimal_trajectory}
\end{equation}
where the superscript $*$ denotes the optimal solution.

The first deployment action is applied to the system
\begin{equation}
u^M_{z,k}
=
u^{M*}_{z,0},
\qquad
a^T_{z,k}
=
a^{T*}_{z,0}.
\label{eq:mpc_first_action}
\end{equation}
The remaining predicted actions are updated when the optimization is solved again at the next control step.

The actual aerial service is calculated using the deployed UAVs, the communication load requiring aerial support, and the available backhaul capacity. The backlog and Mobile UAV energy states are then updated to the next control step according to the system dynamics defined in \Cref{sec:math_dynamics}. The process is repeated at each control step, allowing UAV deployment to adapt as system conditions change.

% This repeated optimization allows the active aerial fleet to adapt as ground recovery, communication load, backlog, and Mobile UAV energy availability evolve over time. The resulting deployment therefore forms a sequence of state dependent fleet sizing decisions rather than a fleet schedule fixed at the beginning of the recovery period.

\subsubsection{Piecewise Affine Relay Approximation}
\label{sec:sol_approx}

The aerial service model in \Cref{eq:aerial_service_capacity} uses the nonlinear saturation function
\begin{equation}
\phi(v)
=
\frac{v}{1+\alpha v}.
\label{eq:relay_saturation_function}
\end{equation}
The function is increasing and concave for $v\ge0$. Since the original expression is nonlinear, we use a conservative piecewise affine approximation in the MPC optimization.

Let
\begin{equation}
v_{\mathrm{focus}} < v_{\max}
\end{equation}
where $v_{\mathrm{focus}}$ defines the range in which more approximation points are placed. A knot is an input value that divides the nonlinear function into intervals for the piecewise affine approximation. The knot locations are defined as
\begin{equation}
\nu_j
=
v_{\mathrm{focus}}
\left(
\frac{j}{J-1}
\right)^{p},
\qquad
j=0,\ldots,J-1,
\label{eq:pwa_inner_knots}
\end{equation}
with the final knot given by
\begin{equation}
\nu_J=v_{\max},
\label{eq:pwa_final_knot}
\end{equation}
where $p>1$ controls the concentration of knots in the lower relay input range. The final knot at $v_{\max}$ ensures that the full relay input range remains covered. For interval $j$, the affine segment connecting $\phi(\nu_{j-1})$ and $\phi(\nu_j)$ is defined by
\begin{equation}
m_j
=
\frac{
\phi(\nu_j)-\phi(\nu_{j-1})
}{
\nu_j-\nu_{j-1}
},
\label{eq:pwa_slope}
\end{equation}
and
\begin{equation}
b_j
=
\phi(\nu_{j-1})
-
m_j\nu_{j-1}.
\label{eq:pwa_intercept}
\end{equation}

The MPC uses the following aerial service constraint
\begin{equation}
y_{z,h}
\le
q_{\mathrm{ref}}
\left(
m_j v_{z,h}+b_j
\right),
\qquad
j=1,\ldots,J,
\qquad
\forall z,h.
\label{eq:pwa_service_constraint}
\end{equation}

Since $\phi(v)$ is concave, the piecewise affine approximation remains below the original nonlinear function over the relay input range. The service limit does not overestimate the nonlinear aerial service capacity.

The approximation is used only within the MPC optimization. After the first deployment action is implemented, the closed loop simulation evaluates the realized aerial service using the nonlinear saturation function in \Cref{eq:relay_saturation_function}. Therefore, the optimization is solved as a convex quadratic program, while the simulated system preserves the original nonlinear service model. The number of affine segments and the corresponding approximation parameters used in the numerical experiments are reported in \Cref{tab:case_study_parameters}.

\section{Numerical Experiments}
\label{sec:numerical_exp}
This section evaluates the proposed MPC framework across historical disaster recovery and backhaul scenarios. We first describe the experimental setup, followed by analyses of dynamic fleet sizing, model components and system bottlenecks, sensitivity and robustness, and computational performance.

\subsection{Experimental Setup}
\label{sec:exp_setup}
The experimental setup defines the historical recovery cases, backhaul scenarios, model parameters, Monte Carlo design, evaluation metrics, and computational implementation used throughout the numerical study.

\subsubsection{Historical Disaster Recovery Cases}
\label{sec:exp_historical}

Seven historical disaster recovery cases are used to represent a broad range of ground communication recovery patterns. The cases include the West Maui wildfire and Hurricanes Maria, Helene, and Michael across multiple affected regions. Historical communication outage and recovery observations are obtained from disaster reporting records, including data reported through the Federal Communications Commission Disaster Information Reporting System \cite{fcc2018AtlanticHurricanes,fcc2019HurricaneMichael,fcc2023HawaiiWildfires,fcc2024HurricaneHelene}. The logistic recovery model introduced in \Cref{sec:sys_recovery_demand} is fitted separately to each disaster and region. \Cref{tab:historical_recovery_cases} summarizes the number of observations and the fitted recovery parameters. Regional abbreviations introduced in the table are used throughout the numerical results.

\begin{table}
\centering
\footnotesize
\renewcommand{\arraystretch}{1.1}
\setlength{\tabcolsep}{3pt}
\begin{tabularx}{\textwidth}{@{}ll>{\raggedright\arraybackslash}Xrrrrr@{}}
\toprule
Type
& Name
& Region
& $n$
& $K$
& \begin{tabular}[c]{@{}c@{}}$r$\\(per hour)\end{tabular}
& \begin{tabular}[c]{@{}c@{}}$\tau$\\(hours)\end{tabular}
& $R^2$ \\
\midrule
Wildfire
& West Maui
& West Maui
& 12
& 0.997
& 0.0365
& 187
& 0.994 \\

Hurricane
& Maria
& Puerto Rico (PR)
& 103
& 0.942
& 0.00222
& 1060
& 0.997 \\

Hurricane
& Maria
& U.S. Virgin Islands (USVI)
& 102
& 0.887
& 0.00124
& 583
& 0.971 \\

Hurricane
& Helene
& North Carolina (NC)
& 22
& 0.931
& 0.0164
& 103
& 0.985 \\

Hurricane
& Helene
& South Carolina (SC)
& 10
& 0.994
& 0.0159
& -49.3
& 0.997 \\

Hurricane
& Helene
& Virginia (VA)
& 6
& 1.00
& 0.0315
& 21.2
& 0.984 \\

Hurricane
& Michael
& Florida, Georgia, Alabama
& 10
& 0.956
& 0.0334
& -27.7
& 0.947 \\
\bottomrule
\end{tabularx}
\caption{Historical disaster cases and fitted ground recovery parameters.}
\label{tab:historical_recovery_cases}
\end{table}
In \Cref{tab:historical_recovery_cases}, $n$ is the number of historical observations used for each fit. The number of observations varies across cases since the available disaster reports provide different numbers of recovery updates. The parameters $K$, $r$, and $\tau$ indicate the maximum recovery level, recovery rate, and the time when the recovery rate is highest. The high $R^2$ values, ranging from 0.947 to 0.997, show that the logistic model fits the observed recovery patterns well. A negative value of $\tau$ indicates that the fastest recovery occurred before $t=0$.

The fitted cases show substantial differences in recovery stage and recovery rate. The Hurricane Maria cases in Puerto Rico and the U.S. Virgin Islands show relatively slow recovery over the evaluation period. The West Maui case also remains at an early recovery stage for much of the study period. Hurricane Helene in North Carolina shows a clear recovery transition within the evaluation period, while the South Carolina and Virginia cases begin at higher recovery levels. Hurricane Michael also represents a relatively advanced recovery case. These differences produce different communication loads requiring aerial support across the disaster cases.

% Parameter uncertainty is summarized using approximate 95\% confidence intervals obtained from the covariance matrix of the nonlinear logistic fit. The interval for each parameter is computed from its fitted standard error using a two-sided Student's $t$ critical value with $n-3$ degrees of freedom. 

% The corresponding confidence intervals for $K$, $r$, and $\tau$ are provided in the Supplementary Material.

\begin{figure}
\centering
\includegraphics[width=0.65\linewidth]{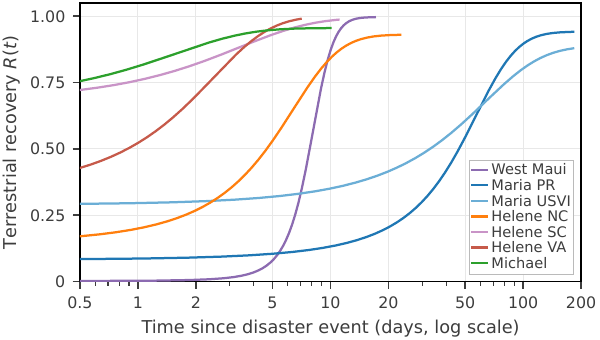}
\caption{Fitted ground recovery trajectories for the seven historical disaster cases.}
\label{fig:historical_recovery}
\end{figure}

The corresponding recovery trajectories are shown in \Cref{fig:historical_recovery}, highlighting the different recovery patterns represented by the seven historical cases.

\Cref{fig:historical_recovery} illustrates the large variation in recovery dynamics represented by the historical cases. Maria PR and Maria USVI show slow recovery over extended periods, while Helene NC transitions more rapidly toward its fitted maximum recovery level. West Maui also shows a delayed recovery transition. Helene SC, Helene VA, and Michael begin at relatively advanced recovery levels and approach their maximum recovery levels more quickly. The time axis is shown on a logarithmic scale from 0.5 days onward to compare recovery patterns that occur over different time scales.

The numerical experiments use the fitted recovery models to generate ground communication capacity over the 168 hours evaluation horizon. Historical outage observations are used to estimate the recovery trajectory $R(t)$ for each disaster case, which determines $G_{z,k}$ through \Cref{eq:ground_capacity}. The communication demand model, baseline ground communication capacities, UAV fleet parameters, and Tethered UAV site locations are specified separately and held fixed across disaster cases in the primary experiment. Each disaster case is evaluated under the same four backhaul scenarios. The experiments therefore compare historical recovery patterns under common demand and operational assumptions.

\subsubsection{Backhaul Scenarios}
\label{sec:exp_backhaul}

\begin{table}
\footnotesize
\centering
\renewcommand{\arraystretch}{1.1}
\begin{tabular}{lcc}
\toprule
Backhaul scenario
& Capacity (Mbps)
& \begin{tabular}[c]{@{}c@{}}
$C_{b}^{\mathrm{bh}}$
(service units per hour)
\end{tabular} \\
\midrule
Fiber          & 1000 & 5.0 \\
Microwave      & 500  & 2.5 \\
Mobile Gateway & 200  & 1.0 \\
Satellite      & 100  & 0.5 \\
\bottomrule
\end{tabular}
\caption{Backhaul capacity scenarios used in the case study.}
\label{tab:backhaul_scenarios}
\end{table}

\Cref{tab:backhaul_scenarios} summarizes the four backhaul scenarios considered in the numerical experiments. We assume that each service zone has its own backhaul capacity, which is not shared across zones. Within each scenario, all six service zones use the same backhaul type and capacity so that the effect of backhaul availability can be evaluated separately. Independent backhaul connections are used in emergency communication systems, where each temporary cell site uses its own backhaul connection, such as a satellite link \cite{firstnetDeployables}.

Representative commercial systems and communication standards were used as references when selecting the backhaul capacities for the case study. The Fiber case uses 1000 Mbps, compared with GPON systems that support downstream rates around 2.5 Gbps \cite{itu2008g9841}. The Microwave case uses 500 Mbps, with the Cambium PTP 820S reporting throughput up to approximately 600 Mbps \cite{cambiumPTP820S}. The Mobile Gateway case uses 200 Mbps, based on the rated secure throughput of the Peplink MAX BR2 Pro \cite{peplinkBR2Pro}. The Satellite case uses 100 Mbps as a representative capacity since actual satellite throughput varies with service conditions such as network load, service plan, and terminal type \cite{starlinkSpecifications}. Satellite links are also used in rapidly deployable communication systems when ground infrastructure is unavailable \cite{firstnetDeployables}.

Using the service scaling factor $k_{\mathrm{eq}}=0.005$, the physical backhaul capacities are mapped directly to
\begin{equation}
C_{b}^{\mathrm{bh}}
=
k_{\mathrm{eq}} Q_{b}^{\mathrm{bh}}.
\end{equation}
The corresponding service capacities are 5.0, 2.5, 1.0, and 0.5 service units per hour for Fiber, Microwave, Mobile Gateway, and Satellite, respectively. 

\subsubsection{Model and Experimental Parameters}
\label{sec:exp_parameters}

\begin{table}
\centering
\footnotesize
\renewcommand{\arraystretch}{1.1}
\setlength{\tabcolsep}{4pt}
\begin{tabular}{lccc}
\toprule
Parameter & Symbol & Value & Unit \\
\midrule

Number of service zones
    & $Z$ & 6 & -- \\

Control time step
    & $\Delta t$ & 1 & hours \\

Prediction horizon
    & $H$ & 8 & steps \\

Evaluation horizon
    & $T_{\mathrm{eval}}$ & 168 & hours \\

Mobile UAV fleet size
    & $N_M$ & 30 & UAVs \\

Tethered UAV fleet size
    & $N_T$ & 3 & UAVs \\

Mobile UAV flight endurance
    & $T_M^{\mathrm{flight}}$ & 1 & hours \\

Mobile UAV recharge time
    & $T_M^{\mathrm{chg}}$ & 1.5 & hours \\

Mobile UAV deployment ramp
    & $R_M$ & 8 & UAVs per hour \\

Mobile UAV service gain
    & $g_M$ & 1.0 & -- \\

Tethered UAV service gain
    & $g_T$ & 1.0 & -- \\

Demand scaling factor
    & $s_D$ & 0.50 & -- \\

Demand-to-service conversion
    & $k_{\mathrm{eq}}$ & 0.005
    & service units/(Mbps$\cdot$hour) \\

Reference communication capacity
    & $Q_{\mathrm{ref}}$ & 100 & Mbps \\

Reference service rate
    & $q_{\mathrm{ref}}$ & 0.5 & service units per hour \\

Service saturation coefficient
    & $\alpha$ & 0.01 & -- \\

Backlog penalty weight
    & $c_x$ & 50 & -- \\

Aerial deployment penalty weight
    & $c_u$ & 0.001 & -- \\

Deployment variation penalty weight
    & $c_s$ & 0.05 & -- \\

Terminal backlog penalty weight
    & $c_T$ & 20 & -- \\

Mobile UAV energy preference weight
    & $\epsilon_E$ & $10^{-6}$ & -- \\

Piecewise affine segments
    & $J$ & 6 & -- \\

PWA full relay input range
    & $v_{\max}$ & 40 & -- \\

PWA focus point
    & $v_{\mathrm{focus}}$ & 20 & -- \\

PWA knot spacing exponent
    & $p$ & $3/2$ & -- \\

\bottomrule
\end{tabular}
\caption{Model and experimental parameters for the case study. A dash indicates a dimensionless parameter or tuning coefficient.}
\label{tab:case_study_parameters}
\end{table}

\Cref{tab:case_study_parameters} shows the baseline model and experimental parameters used throughout the case study. Since the control time step is 1 hour, the 8 steps prediction horizon corresponds to 8 hours. The demand-to-service conversion factor $k_{\mathrm{eq}}$ converts communication capacity in Mbps to service units per hour. The UAV fleet consists of 30 Mobile UAVs and three Tethered UAVs serving six demand zones over the 168 hours evaluation period. The Mobile UAV flight endurance is set to 1 hour as a conservative value based on currently available UAV platforms. For example, the PDW C100 reports a maximum flight time of approximately 74 min with an ISR payload and supports communication payload integration \cite{pdwC100}. The Tethered UAV model is based on the long operating time enabled by continuous ground power. For reference, the Elistair Khronos supports continuous operation and uses a tether that can also carry data for communication payloads \cite{elistairKhronos}. 

A common nominal communication demand profile is used across all disaster and backhaul scenarios. Demand varies across service zones and over time through daily and shorter period variations together with temporary demand surges. The baseline demand scale is $s_D=0.50$, and the same nominal profile is used for all primary Monte Carlo comparisons so that differences among cases are driven by recovery conditions, backhaul capacity, and the tested model components.
The baseline ground communication capacity in each zone is set to 105\% of the maximum nominal demand in that zone and remains fixed across the primary experiments. The three Tethered UAV sites are assigned to the zones with the highest mean nominal demand and remain fixed throughout the evaluation.
Each simulation begins with zero communication backlog and no active UAV deployment. Models with explicit Mobile UAV energy dynamics start with the fleet fully charged.

% Several parameters in \Cref{tab:case_study_parameters} are selected for the numerical experiments rather than taken directly from hardware specifications. These include the recharge time, fleet sizes, deployment ramp limit, prediction horizon, and demand scaling factor. The recharge time and fleet composition are varied in separate sensitivity analyses. The service gains and service scaling factor define the communication model and do not represent the performance of a specific UAV or radio system. For each disaster and backhaul scenario, 30 matched Monte Carlo realizations are used to compare the model variants under the same conditions.

\subsubsection{Monte Carlo Experiment Setup}
\label{sec:exp_MC_design}

We evaluate six model cases in the Monte Carlo study. The \textit{Integrated} model includes ground recovery information, backhaul constraints, the mixed Mobile and Tethered UAV fleet, and explicit Mobile UAV energy dynamics. Five ablation cases are used to examine the effect of each model component. \textit{Persistence} assumes that the current ground recovery level remains unchanged over the prediction horizon. \textit{No Recovery} removes ground recovery information from the planning process. \textit{No Backhaul} ignores the backhaul capacity during optimization, while the actual backhaul limit is still applied in the simulation. \textit{All Mobile} replaces the three Tethered UAVs with three additional Mobile UAVs, resulting in a fleet of 33 Mobile UAVs. \textit{Duty Limit} replaces the Mobile UAV energy model with a fixed availability limit based on the flight and recharge times.

For a consistent comparison, all six model cases are evaluated using the same random realization for each scenario. Seven historical disaster cases are combined with four backhaul scenarios and 30 Monte Carlo runs for each disaster and backhaul combination, resulting in 840 matched scenarios. Each of the six model cases is evaluated for all 840 scenarios. \Cref{tab:mc_design} summarizes the experiment matrix.

\begin{table}
\centering
\footnotesize
\label{tab:mc_design}
\renewcommand{\arraystretch}{1.1}
\begin{tabular}{lr}
\toprule
Experiment setting & Value \\
\midrule
Historical disaster cases & 7 \\
Backhaul scenarios & 4 \\
Monte Carlo samples for each disaster and backhaul case & 30 \\
Total matched scenarios & 840 \\
Model cases tested in each scenario & 6 \\
Total simulation runs & 5,040 \\
Comparisons with the Integrated model & 4,200 \\
\bottomrule
\end{tabular}
\caption{Monte Carlo experiment setup.}
\end{table}

The same Monte Carlo samples are used for the Integrated model and all five ablation cases, allowing direct comparisons under the same conditions. The 840 scenarios are evaluated for all six model cases, resulting in 5,040 simulation runs. Comparing the Integrated model with each of the five ablation cases across these scenarios gives 4,200 comparisons.

\subsubsection{Evaluation Metrics}
\label{sec:exp_metrics}

System performance is evaluated using service, backlog, and aerial resource metrics over the 168 hours evaluation horizon. The served ratio measures the cumulative communication service delivered over the evaluation horizon relative to the cumulative communication demand and is reported as a percentage. The backlog represents communication demand that is temporarily deferred when available service is insufficient at a given time, and the backlog integral measures the accumulated backlog over time by integrating the queue over the evaluation period. Backlogged demand can be served later when sufficient capacity becomes available. In other words, a 100\% served ratio is still possible even when backlog occurs temporarily, as long as all delayed demand is served later within the evaluation period. 
Finally, the served ratio captures the total amount of demand ultimately served, while the backlog integral indicates the duration and magnitude of temporary service delay.

Aerial resource use is measured by total relay hours, which is the total operating time of active Mobile UAVs and Tethered UAVs across all service zones. This metric shows how much UAV support is used to provide the reported service level. Moreover, the trajectory analysis examines Mobile UAV utilization and fleet energy to show how deployment and energy availability change over time.

For each Monte Carlo sample, we combine the backlog across all disaster and backhaul cases to calculate the backlog reduction between models. Let $B_{d,b,m}^{(a)}$ denote the backlog integral for ablation case $a$ under disaster $d$, backhaul scenario $b$, and Monte Carlo sample $m$. The corresponding backlog reduction of the Integrated model is defined as
\begin{equation}
\Delta B_m^{(a)}
=
100
\left(
\frac{
\sum_{d}\sum_{b} B_{d,b,m}^{(a)}
-
\sum_{d}\sum_{b} B_{d,b,m}^{(\mathrm{Integrated})}
}{
\sum_{d}\sum_{b} B_{d,b,m}^{(a)}
}
\right),
\label{eq:pooled_backlog_reduction}
\end{equation}
where factor of 100 converts the relative backlog reduction into a percentage. The overall backlog reduction is the average of $\Delta B_m^{(a)}$ across the $N_{\mathrm{MC}}=30$ Monte Carlo samples. Combining the backlog across disaster and backhaul cases avoids large relative changes caused by cases with very small backlog. Positive values mean that the Integrated model has lower backlog than the ablation case. Served ratio differences are reported in percentage points, while relay hour differences show changes in UAV use. The prediction horizon sensitivity analysis also reports the percentage of control steps with different fleet decisions. The same Monte Carlo samples are used for each model comparison to separate the effect of model changes from differences in random conditions.

\subsubsection{Computational Implementation}
\label{sec:exp_implementation}

The MPC problems are implemented in Python using CVXPY 1.9.2 and solved as convex quadratic programs. CLARABEL is used as the main solver, while OSQP and SCS are used as fallback solvers when needed. Solver outputs are checked for constraint violations before they are used in the simulation. 

% Solver outputs are independently checked for constraint violations before the corresponding control action is implemented. Candidate solutions that do not satisfy the scale-aware numerical tolerance are rejected, and the next solver in the fallback sequence is attempted. After a solution passes this residual check, the first-step continuous fleet allocation is projected onto the applicable deployment, availability, and site constraints to remove small solver-level numerical violations. This projection is only a numerical feasibility correction and does not round the continuous fleet decisions to integer values; solutions requiring a correction greater than \(10^{-3}\) UAV are rejected.

The simulation updates the UAV deployment, communication service, backlog, and Mobile UAV energy at each control step using the optimized MPC decision. All numerical experiments were performed on a computer with a 13th Gen Intel Core i7-1360P processor and 32 GB of LPDDR5 memory. To reduce computation time, the Monte Carlo experiments were run using eight parallel workers.

\subsection{Dynamic Fleet Sizing Across Disasters}
\label{sec:exp_main_result}
This subsection examines how communication performance, active fleet deployment, backlog, and Mobile UAV energy evolve across different disaster recovery conditions.

\subsubsection{System Performance}
\label{sec:exp_main_performance}

We examine the performance of the Integrated model across the historical disaster and backhaul scenarios. \Cref{tab:integrated_system_performance} reports the served ratio, backlog integral, and total relay hours for each combination of disaster and backhaul condition. The results are averaged across 30 Monte Carlo samples using the performance metrics defined in \Cref{sec:exp_metrics}.

\begin{table}
\centering
\footnotesize
\renewcommand{\arraystretch}{1.08}
\setlength{\tabcolsep}{8pt}
\begin{tabular}{@{}llccc@{}}
\toprule
Disaster
& Backhaul
& \begin{tabular}[c]{@{}c@{}}Served ratio\\(\%)\end{tabular}
& \begin{tabular}[c]{@{}c@{}}Backlog integral\\(service units $\cdot$ hour)\end{tabular}
& \begin{tabular}[c]{@{}c@{}}Relay hours\\(UAV hour)\end{tabular} \\
\midrule

\multirow{4}{*}{West Maui}
& Fiber          & $68.8 \pm 0.2$ & $56{,}460 \pm 490$ & $2{,}530.8$ \\
& Microwave      & $68.8 \pm 0.2$ & $56{,}470 \pm 490$ & $2{,}530.8$ \\
& Mobile Gateway & $56.4 \pm 0.2$ & $75{,}630 \pm 490$ & $2{,}083.6 \pm 1.6$ \\
& Satellite      & $28.2 \pm 0.1$ & $118{,}180 \pm 490$ & $1{,}074.1 \pm 6.0$ \\
\midrule

\multirow{4}{*}{Maria PR}
& Fiber          & $72.9 \pm 0.3$ & $39{,}080 \pm 490$ & $2{,}530.8$ \\
& Microwave      & $72.9 \pm 0.3$ & $39{,}080 \pm 490$ & $2{,}530.8$ \\
& Mobile Gateway & $59.6 \pm 0.2$ & $58{,}490 \pm 480$ & $2{,}074.6 \pm 0.7$ \\
& Satellite      & $29.9 \pm 0.1$ & $100{,}790 \pm 490$ & $1{,}058.9 \pm 5.0$ \\
\midrule

\multirow{4}{*}{Maria USVI}
& Fiber          & 100.0 & $12.8 \pm 3.5$ & $2{,}502.3 \pm 2.5$ \\
& Microwave      & 100.0 & $12.8 \pm 3.5$ & $2{,}502.3 \pm 2.5$ \\
& Mobile Gateway & $91.2 \pm 0.4$ & $9{,}910 \pm 400$ & $2{,}031.1 \pm 2.0$ \\
& Satellite      & $47.6 \pm 0.3$ & $48{,}110 \pm 480$ & $1{,}027.6 \pm 0.7$ \\
\midrule

\multirow{4}{*}{Helene NC}
& Fiber          & 100.0 & $1{,}809 \pm 99$ & $1{,}990.6 \pm 5.8$ \\
& Microwave      & 100.0 & $1{,}809 \pm 99$ & $1{,}990.6 \pm 5.8$ \\
& Mobile Gateway & $99.0 \pm 0.3$ & $11{,}510 \pm 330$ & $1{,}837.2 \pm 8.8$ \\
& Satellite      & $59.7 \pm 0.4$ & $48{,}730 \pm 470$ & $1{,}028.1 \pm 1.1$ \\
\midrule

\multirow{4}{*}{Helene SC}
& Fiber          & 100.0 & $<0.01$ & $197.4 \pm 6.4$ \\
& Microwave      & 100.0 & $<0.01$ & $197.4 \pm 6.4$ \\
& Mobile Gateway & 100.0 & $<0.01$ & $197.4 \pm 6.4$ \\
& Satellite      & 100.0 & $5.6 \pm 1.9$ & $195.8 \pm 6.3$ \\
\midrule

\multirow{4}{*}{Helene VA}
& Fiber          & 100.0 & $<0.01$ & $544.0 \pm 7.1$ \\
& Microwave      & 100.0 & $<0.01$ & $544.0 \pm 7.1$ \\
& Mobile Gateway & 100.0 & $139.5 \pm 6.4$ & $530.1 \pm 6.5$ \\
& Satellite      & 100.0 & $1{,}546 \pm 58$ & $422.9 \pm 4.2$ \\
\midrule

\multirow{4}{*}{Michael}
& Fiber          & 100.0 & $<0.01$ & $112.1 \pm 5.2$ \\
& Microwave      & 100.0 & $<0.01$ & $112.1 \pm 5.2$ \\
& Mobile Gateway & 100.0 & $<0.01$ & $112.1 \pm 5.2$ \\
& Satellite      & 100.0 & $1.16 \pm 0.53$ & $111.4 \pm 5.3$ \\
\bottomrule
\end{tabular}
\caption{System performance of the Integrated model across historical disaster and backhaul scenarios.}
\label{tab:integrated_system_performance}
\end{table}

The values in \Cref{tab:integrated_system_performance} are reported as mean plus or minus one standard deviation across 30 Monte Carlo samples. Standard deviations smaller than the displayed precision are omitted. Backlog values below 0.01 service units $\cdot$ hours are reported as $<0.01$ without standard deviation.

The results show that system performance depends strongly on both the ground recovery trajectory and the available backhaul capacity. Fiber and Microwave provide high backhaul capacity and produce nearly the same performance across all seven disaster cases. When the communication load requiring aerial support remains high, lower backhaul capacity leads to lower system performance. For West Maui, the served ratio decreases from 68.8\% under Fiber to 28.2\% under Satellite, while the backlog integral increases from 56,460 to 118,180 service units $\cdot$ hours. Maria PR shows a similar pattern, with the served ratio decreasing from 72.9\% to 29.9\%.

The effect of backhaul capacity becomes smaller as ground recovery reduces the residual communication load. Maria USVI achieves a served ratio of 100\% under Fiber and Microwave, while performance decreases under Mobile Gateway and Satellite. Helene NC also serves all demand under Fiber and Microwave. However, the positive backlog integral shows that some demand is temporarily delayed and served later as the ground network recovers. In contrast, Helene SC, Helene VA, and Michael maintain a served ratio of approximately 100\% across all four backhaul scenarios. Their smaller backlog values show that the recovering ground network leaves less demand for the aerial relay system.

Furthermore, relay use varies across different disaster and backhaul conditions. West Maui and Maria PR use more than 2,500 relay hours under Fiber and Microwave, showing high reliance on aerial support. Relay use decreases as the backhaul becomes more restrictive, since additional aerial capacity cannot always be translated into delivered communication service. The difference becomes larger under Satellite, where limited backhaul capacity reduces the benefit of using additional UAVs. In contrast, Michael uses only about 112 relay hours across the tested backhaul conditions, while Helene SC uses about 197 relay hours. In these cases, ground recovery reduces the amount of aerial support needed over time.

\begin{figure}
\centering
\includegraphics[width=\linewidth]{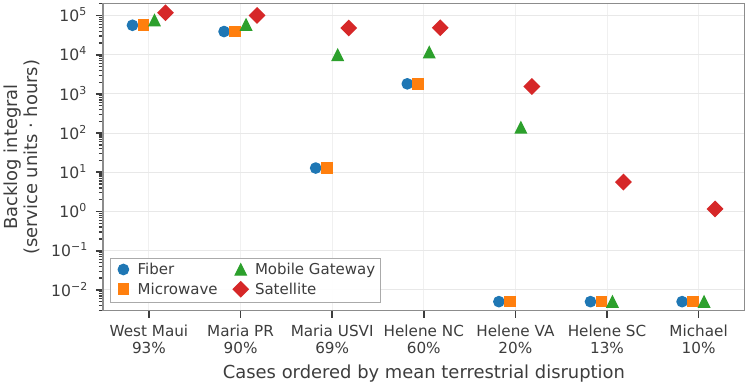}
\caption{Backlog integral across historical disaster cases and backhaul scenarios.}
\label{fig:recovery_backhaul_regime}
\end{figure}

The results show three different system behavior patterns under the same case study setup. Some disasters experience continued communication shortages, while others return to full service as the ground network recovers. In other cases, only limited aerial support is needed for most of the evaluation period. To further relate these differences to ground recovery severity, we define the mean ground disruption over the evaluation horizon as
\begin{equation}
\bar{D}^{\mathrm{rec}} =
\frac{1}{T_{\mathrm{eval}}}
\int_{0}^{T_{\mathrm{eval}}}
\left(1-R(t)\right)\,dt,
\label{eq:mean_ground_disruption}
\end{equation}
where $R(t)$ denotes the fitted ground recovery trajectory and $T_{\mathrm{eval}}=168$ hours. The quantity $\bar{D}^{\mathrm{rec}}$ represents the average fraction of ground service remaining disrupted over the evaluation horizon. Larger values indicate slower or more persistent ground recovery. For clarity, \Cref{fig:recovery_backhaul_regime} shows this value as a percentage and arranges the disaster cases by average ground disruption.

\Cref{fig:recovery_backhaul_regime} compares the backlog across the historical disaster cases and backhaul scenarios. The disaster cases are ordered from higher to lower mean ground disruption, with the disruption percentage shown below each case. A logarithmic scale is used for the backlog since the values vary widely across the cases. Backlog values below $0.01$ service units $\cdot$ hours are plotted at $0.005$ service units $\cdot$ hours for visualization.

The result shows that system burden depends on both ground recovery and backhaul capacity. Cases with more persistent ground disruption, including West Maui and Maria PR, maintain large backlog levels across all backhaul scenarios. As ground disruption decreases, backlog generally falls, while the available backhaul capacity affects how much backlog remains. Notably, Fiber and Microwave result in low backlog when ground recovery is more advanced, while Mobile Gateway and Satellite lead to higher backlog. This pattern shows that ground recovery alone does not determine system performance. The same recovery trajectory can lead to different backlog levels when the available backhaul capacity changes.

\subsubsection{Service, Backlog, and Fleet Trajectories}
\label{sec:exp_main_fleet_traj}

\begin{figure}[t]
\centering
\includegraphics[width=\linewidth]{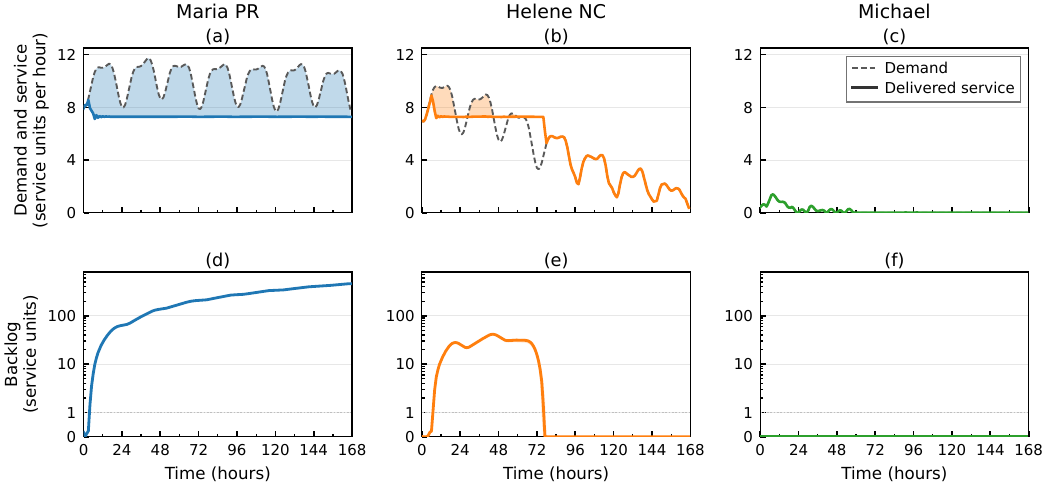}
\caption{Communication service and backlog trajectories for Maria PR, Helene NC, and Michael under the Integrated model with Fiber backhaul.}
\label{fig:service_backlog_trajectories}
\end{figure}

\begin{figure}[t]
\centering
\includegraphics[width=\linewidth]{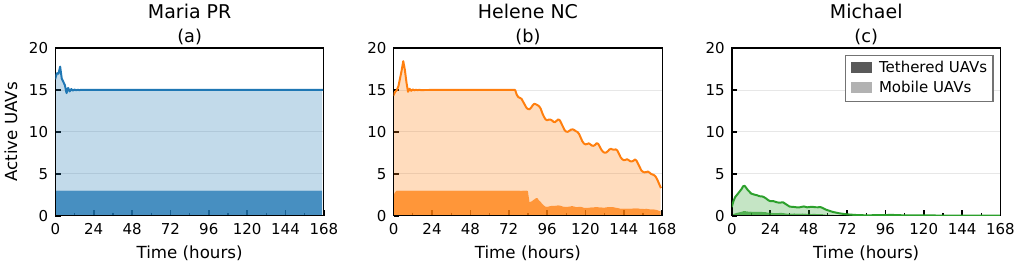}
\caption{Active UAV fleet composition for Maria PR, Helene NC, and Michael under the Integrated model with Fiber backhaul.}
\label{fig:fleet_trajectories}
\end{figure}

To examine how terrestrial recovery translates into communication load and aerial resource use over time, we next consider three representative disaster cases under the Integrated model with Fiber backhaul. Maria PR represents persistent communication shortage, Helene NC shows a clear recovery transition within the evaluation period, and Michael represents a low workload case after the terrestrial network has largely recovered. The corresponding terrestrial recovery trajectories are shown in \Cref{fig:historical_recovery}. \Cref{fig:service_backlog_trajectories} compares communication demand, delivered service, and backlog for the three representative cases, while \Cref{fig:fleet_trajectories} shows the corresponding active UAV fleet composition. The trajectories report the median across 30 Monte Carlo realizations. In the service panels, the shaded area between communication demand and delivered service indicates unmet demand. In the backlog panels, the shaded bands indicate the 10th--90th percentile range across realizations. A symmetric logarithmic scale is used for the backlog because the backlog levels vary widely across the three cases. In the fleet panels, the stacked areas show the median numbers of active Tethered UAVs and Mobile UAVs.

The three cases exhibit distinctly different operating patterns. Maria PR remains at a low terrestrial recovery level throughout the evaluation horizon. Communication demand exceeds delivered service repeatedly over the evaluation period, resulting in persistent unmet demand and sustained aerial relay use. The median backlog continues to accumulate throughout the horizon and reaches more than 450 service units near the end of the period. The active fleet remains relatively stable for most of the evaluation period, showing continued need for aerial support.

Helene NC shows a clear shift from high aerial support to lower aerial support as the terrestrial network recovers. The median backlog initially increases and reaches a peak of 41.4 service units at 44 hours. As terrestrial recovery progresses, residual communication demand decreases and the backlog gradually clears. The aerial fleet responds to the same transition. Mobile UAV deployment decreases substantially after the period of temporary congestion, while Tethered UAV use also declines as less relay capacity is required. By the end of the evaluation, the median number of active Mobile UAVs decreases to 2.9. Dynamic fleet sizing therefore reduces the number of active UAVs as terrestrial communication capacity recovers.

Michael represents a case with low communication demand for aerial support. Terrestrial recovery is already high at the beginning of the evaluation period and continues to approach its fitted maximum recovery level. Residual communication demand is small, and both backlog and aerial deployment remain low. Across these cases, the same fleet and demand scaling assumptions lead to different levels of UAV use depending on the terrestrial recovery trajectory.

\subsubsection{Energy Trajectories}
\label{sec:exp_main_energy_traj}

\begin{figure}
\centering
\includegraphics[width=\linewidth]{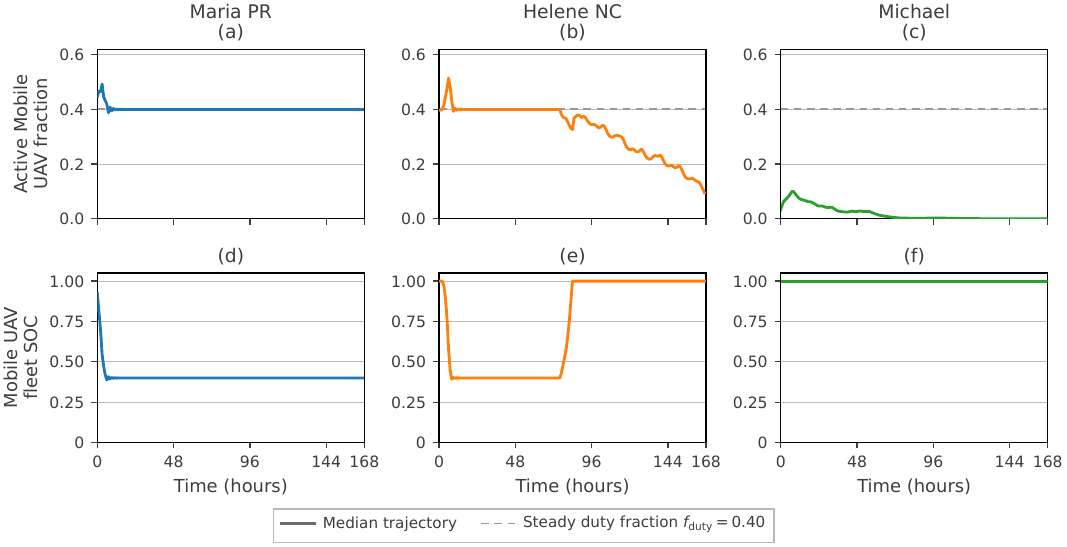}
\caption{Mobile UAV utilization and energy trajectories for Maria PR, Helene NC, and Michael under the Integrated model with Fiber backhaul.}
\label{fig:mobile_energy_trajectories}
\end{figure}

To examine how these fleet deployment patterns interact with Mobile UAV energy availability, \Cref{fig:mobile_energy_trajectories} compares Mobile UAV utilization and fleet state of charge for the same three representative cases. The upper panels show the median fraction of active Mobile UAVs, with the steady duty fraction $f_{\mathrm{duty}}=0.40$ shown by the dashed line. The lower panels show the median Mobile UAV fleet state of charge. All trajectories are computed across 30 Monte Carlo samples.

Maria PR keeps Mobile UAV utilization close to 40\% for most of the evaluation period. This indicates that the Mobile UAV fleet is operating close to the level that can be maintained while UAVs alternate between flying and recharging. The Mobile UAV fleet energy drops quickly from its initial level and stays low since the UAVs continue to be used for aerial support. Helene NC starts with a similar level of Mobile UAV use, and Mobile UAV deployment decreases as the ground network recovers and less aerial support is needed. The fleet state of charge returns toward full availability as fewer Mobile UAVs remain active. Michael shows low Mobile UAV use, and the fleet energy remains close to full for most of the evaluation period.

\begin{figure}
\centering
\includegraphics[width=\linewidth]{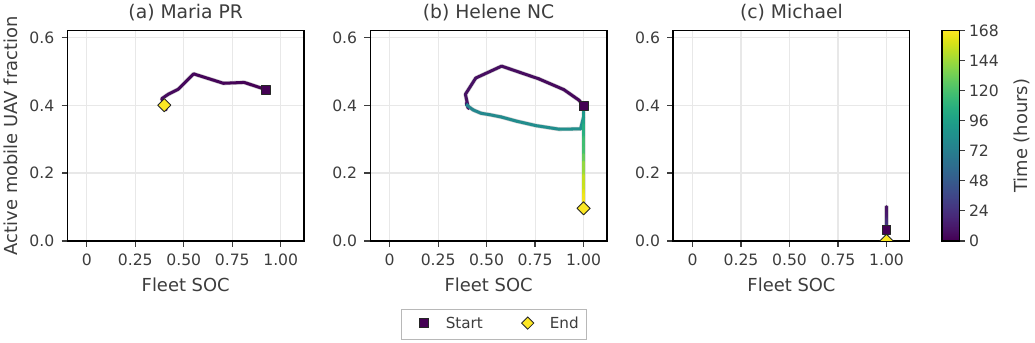}
\caption{Mobile UAV utilization and fleet energy trajectories for Maria PR, Helene NC, and Michael under the Integrated model with Fiber backhaul.}
\label{fig:mobile_energy_phase}
\end{figure}

The previous trajectories show when fleet energy changes over time, while the relationship between Mobile UAV use and fleet energy shows how deployment affects energy availability. \Cref{fig:mobile_energy_phase}  shows the same three cases using the active Mobile UAV fraction and fleet state of charge. Each trajectory is averaged across 30 Monte Carlo samples. The color shows how the system changes over time, while the square and diamond markers indicate the initial and final states.

The phase trajectories in \Cref{fig:mobile_energy_phase} further distinguish the three energy patterns. Maria PR stays near a stable operating condition, with a moderate fleet energy level and about 40\% of the Mobile UAVs active. Helene NC follows a broader trajectory as the system transitions from sustained aerial support toward lower Mobile UAV utilization and renewed energy availability. Michael remains near the high state of charge and low utilization region, consistent with its small residual communication load.

\subsection{Component Ablation and Bottleneck Analysis}
\label{sec:exp_comp_result}
This subsection evaluates the contribution of individual model components and examines how their effects change with backhaul capacity.

\subsubsection{Component Ablation Results}
\label{sec:exp_comp_ablation}

To evaluate the effect of each model component, the Integrated model is compared with the five ablation cases defined in \Cref{sec:exp_MC_design}.  \Cref{tab:component_ablation}  summarizes the results across all historical disaster cases and backhaul scenarios. The values are reported as mean $\pm$ one standard deviation across 30 Monte Carlo samples. For each sample, the seven disaster cases and four backhaul scenarios are combined before the comparison is calculated. Values are rounded to the displayed precision and positive values indicate better performance under the Integrated model.

\begin{table}
\centering
\footnotesize
\renewcommand{\arraystretch}{1.1}
\begin{tabular}{lccc}
\toprule
Ablation case
& Backlog (\%)
& Served ratio (pp)
& Relay hours (UAV hour) \\
\midrule
Persistence & $0.0 \pm 0.0$ & $0.0 \pm 0.0$ & $4.6 \pm 0.2$ \\
No Recovery & $0.2 \pm 0.0$ & $0.0 \pm 0.0$ & $806.6 \pm 3.3$ \\
No Backhaul & $12.0 \pm 0.3$ & $2.6 \pm 0.1$ & $320.9 \pm 1.0$ \\
All Mobile & $7.9 \pm 0.0$ & $1.2 \pm 0.0$ & $-70.6 \pm 0.5$ \\
Duty Limit & $0.6 \pm 0.0$ & $0.0 \pm 0.0$ & $-5.9 \pm 0.9$ \\
\bottomrule
\end{tabular}
\caption{Comparison of the Integrated model with the five ablation cases.}
\label{tab:component_ablation}
\end{table}

The results in \Cref{tab:component_ablation}  show that the model components affect system performance in different ways. Backhaul information has the largest effect. Compared with No Backhaul, the Integrated model reduces backlog by 12.0\%, increases the served ratio by 2.6 percentage points, and uses about 321 fewer relay hours. The fleet composition also has a clear effect. Compared with All Mobile, the Integrated model reduces backlog by 7.9\% and increases the served ratio by 1.2 percentage points. The negative relay hour difference means that the mixed Mobile and Tethered fleet uses about 71 more relay hours while providing better service.

Explicit Mobile UAV energy dynamics have a smaller but measurable effect on service performance. The Integrated model reduces backlog by 0.6\% relative to Duty Limit and increases the served ratio by 0.04 percentage points. The Integrated model uses 5.9 more relay hours. Explicit energy state tracking improves the timing of Mobile UAV deployment and helps maintain deployment feasibility, without simply reducing total aerial resource use.

Ground recovery information has a different effect. The Integrated model reduces the backlog by only 0.2\% relative to No Recovery, but it saves about 800 relay hours. Using current ground recovery information mainly helps avoid unnecessary UAV deployment, with little change in the total service provided. The difference between the Integrated model and Persistence is much smaller. Their backlog and served ratio results are nearly identical at the baseline prediction horizon, while the Integrated model uses 4.6 fewer relay hours. Current ground recovery information helps allocate UAVs more effectively, while future recovery information has little additional effect on backlog over the baseline prediction horizon.

Across the tested scenarios, backhaul information has the largest effect on service performance. Fleet composition and Mobile UAV energy modeling affect performance as well. Ground recovery information has a smaller effect on backlog and helps reduce unnecessary UAV use. Future recovery information provides little additional reduction in backlog at the baseline prediction horizon.

\subsubsection{Effect of Backhaul Capacity}
\label{sec:exp_comp_backhaul}
\begin{figure}
\centering
\includegraphics[width=0.75\linewidth]{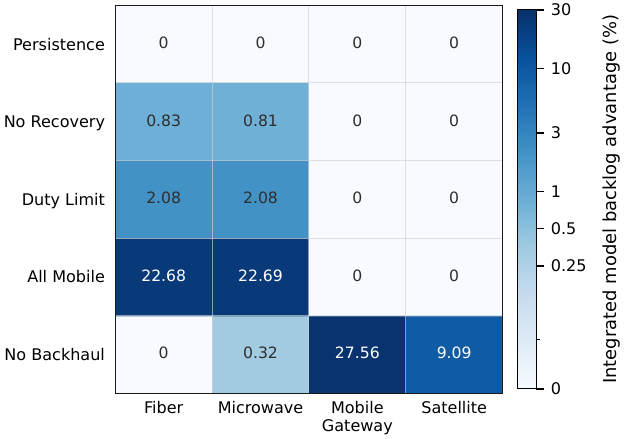}
\caption{Effect of backhaul capacity on backlog performance.}
\label{fig:backhaul_component_effects}
\end{figure}

The overall ablation results do not show how the effects change across backhaul scenarios. \Cref{fig:backhaul_component_effects} shows how the effect of each model component on backlog changes with backhaul capacity. Each value shows the backlog difference between the Integrated model and the corresponding ablation case. Larger positive values mean a greater backlog reduction under the Integrated model.

Under Fiber and Microwave, the All Mobile case with the same total fleet size shows the largest increase in backlog. The Integrated model reduces backlog by 22.7\%, while the effects of Duty Limit and No Recovery are much smaller. Mobile UAV energy dynamics have a smaller effect on backlog. The results show that fleet composition and the use of Tethered UAVs have a greater effect when backhaul capacity is high.

The pattern changes under Mobile Gateway and Satellite. Removing backhaul information has the largest effect. Relative to No Backhaul, the Integrated model reduces backlog by 27.6\% under Mobile Gateway and 9.1\% under Satellite. The effects of fleet composition, Mobile UAV energy dynamics, and ground recovery information are smaller. Additional UAV capacity provides limited benefit when the backhaul cannot support the extra relay traffic. Backhaul capacity and the use of backhaul information have a greater effect under these conditions.

The same difference appears in aerial resource use. When backhaul capacity is low, the system uses fewer relay hours in high demand cases since additional UAVs do not always increase the delivered communication service. When backhaul capacity is high, the system uses more of the available UAV capacity, and fleet composition and Mobile UAV energy have a greater effect on performance. The system performance depends more on UAV resources when backhaul capacity is high and more on backhaul capacity when it is limited.

\subsection{Sensitivity and Robustness Analysis}
\label{sec:exp_sens_result}
This subsection examines the sensitivity of the results to fleet architecture, energy and demand conditions, prediction horizon, and forecast errors.

\subsubsection{Fleet Architecture}
\label{sec:exp_sens_fleet}

The fleet architecture analysis compares the mixed Mobile and Tethered UAV fleet with All Mobile alternatives under two sizing assumptions. The first comparison preserves the total fleet count, while the second matches steady aerial availability based on the Mobile UAV duty cycle. \Cref{tab:fleet_architecture} summarizes the backlog results across the tested scenarios.

\begin{table}
\centering
\footnotesize
\renewcommand{\arraystretch}{1.1}
\setlength{\tabcolsep}{3pt}
\begin{tabular}{@{}>{\raggedright\arraybackslash}p{0.22\textwidth}l>{\centering\arraybackslash}p{0.12\textwidth}>{\centering\arraybackslash}p{0.12\textwidth}>{\centering\arraybackslash}p{0.20\textwidth}@{}}
\toprule
Comparison
& Fleet
& Mobile UAVs
& Tethered UAVs
& Backlog difference (\%) \\
\midrule
\multirow{2}{0.22\textwidth}{Equal count}
& Mixed      & 30 & 3 & 0.00 \\
& All Mobile & 33 & 0 & 7.88 \\
\midrule
\multirow{2}{0.22\textwidth}{Availability matched}
& Mixed      & 30 & 3 & 0.00 \\
& All Mobile & 38 & 0 & -1.06 \\
\bottomrule
\end{tabular}
\caption{Fleet architecture comparison under alternative fleet sizing assumptions.}
\label{tab:fleet_architecture}
\end{table}

Backlog differences in \Cref{tab:fleet_architecture} are reported relative to the mixed fleet. Positive values indicate higher backlog for the All Mobile fleet, while negative values indicate lower backlog. With the same total fleet size, the All Mobile fleet produces 7.9\% higher backlog than the mixed fleet. This difference becomes much smaller when steady aerial availability is matched. The fleet with 38 Mobile UAVs produces 1.1\% lower backlog than the mixed fleet. These results show that the main benefit of Tethered UAVs comes from their longer operating time and efficient use of the available fleet.

\subsubsection{Energy and Demand Scale}
\label{sec:exp_sens_energy}

\begin{figure}
\centering
\includegraphics[width=\linewidth]{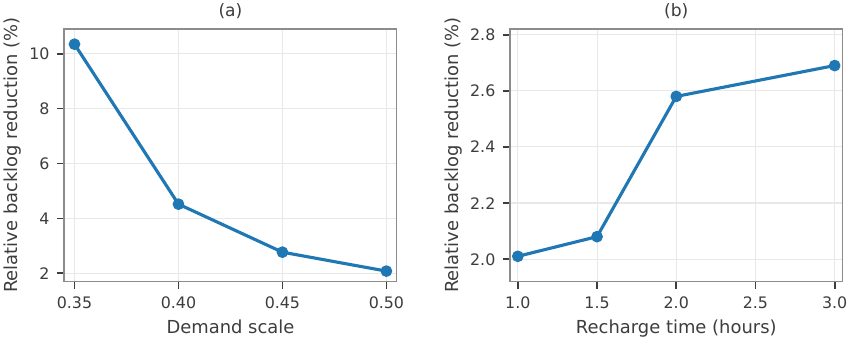}
\caption{Sensitivity of backlog performance to demand scale and Mobile UAV recharge time. (a) shows demand scale and (b) shows recharge time.}
\label{fig:energy_demand_sensitivity}
\end{figure}

We examine how Mobile UAV energy modeling affects performance as communication demand and recharge time change. The Integrated model is compared with Duty Limit using four historical disaster cases, Fiber and Microwave backhaul, and 30 Monte Carlo samples. The demand analysis varies the demand scaling factor from 0.35 to 0.50. The recharge analysis varies the Mobile UAV recharge time from 1 to 3 hours and updates the availability limit used by Duty Limit. \Cref{fig:energy_demand_sensitivity} shows the relative backlog reduction across the tested settings. \Cref{fig:energy_demand_sensitivity}(a) shows the effect of demand scale, while \Cref{fig:energy_demand_sensitivity}(b) shows the effect of Mobile UAV recharge time. Positive values indicate lower backlog under the Integrated model. \Cref{tab:energy_sensitivity} reports both the relative and absolute reductions in backlog achieved by the Integrated model.

\begin{table}
\centering
\footnotesize
\renewcommand{\arraystretch}{1.1}
\begin{tabular}{lrr}
\toprule
Parameter
& \begin{tabular}[c]{@{}c@{}}Relative backlog reduction \\ (\%)\end{tabular}
& \begin{tabular}[c]{@{}c@{}}Absolute backlog reduction \\(service units $\cdot$ hours)\end{tabular} \\
\midrule
\multicolumn{3}{l}{\textit{Demand scale}} \\
\midrule
0.35 & 10.35 & 248 \\
0.40 & 4.52  & 414 \\
0.45 & 2.77  & 465 \\
0.50 & 2.08  & 516 \\
\midrule
\multicolumn{3}{l}{\textit{Mobile UAV recharge time (hours)}} \\
\midrule
1.0 & 2.01 & 288 \\
1.5 & 2.08 & 516 \\
2.0 & 2.58 & 859 \\
3.0 & 2.69 & 1308 \\
\bottomrule
\end{tabular}
\caption{Sensitivity of explicit Mobile UAV energy modeling to demand scale and recharge time.}
\label{tab:energy_sensitivity}
\end{table}

The demand scale results show that the relative benefit of explicit energy modeling decreases as system load increases. In \Cref{fig:energy_demand_sensitivity}(a), the Integrated model reduces backlog by 10.4\% at a demand scale of 0.35, compared with 2.1\% at the baseline value of 0.50. The absolute backlog reduction follows the opposite trend and increases from 248 to 516 service units $\cdot$ hours. Explicit energy state information continues to improve service under higher demand, while its relative effect becomes smaller as overall backlog increases due to the limited system capacity.

The recharge time analysis shows a different pattern. In \Cref{fig:energy_demand_sensitivity}(b), the backlog benefit of the Integrated model increases as Mobile UAV recharge becomes slower. Relative backlog reduction increases from 2.0\% at a 1 hour recharge time to 2.7\% at 3 hours. The absolute reduction in backlog increases from 288 to 1,308 service units $\cdot$ hours. The importance of explicit energy state tracking increases as the interaction between flight and recharge becomes stronger.

The sensitivity results show that the effect of Mobile UAV energy modeling changes with demand and recharge time. The reduction in backlog remains as demand increases and becomes larger when recharge takes longer. A dynamic energy state is useful when Mobile UAV availability changes over time.

\subsubsection{Planning Horizon}
\label{sec:exp_sens_horizon}

\begin{figure}
\centering
\includegraphics[width=\linewidth]{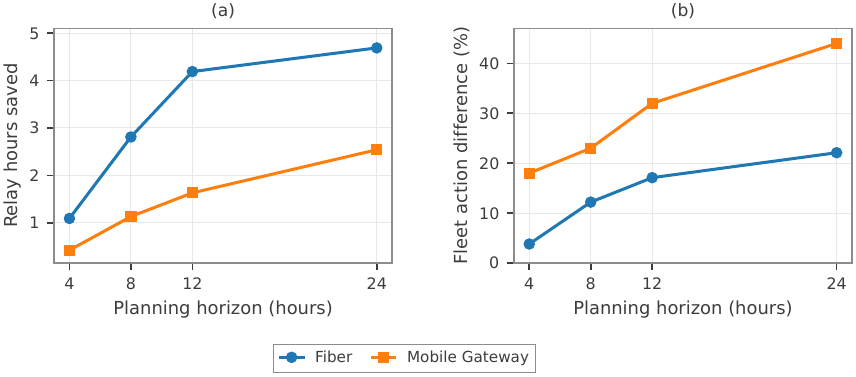}
\caption{Sensitivity of fleet operation to the planning horizon. (a) shows relay hour savings, and (b) shows fleet action differences.}
\label{fig:planning_horizon_sensitivity}
\end{figure}

The planning horizon analysis shows how the value of future ground recovery information changes with the length of the prediction horizon. The Integrated model is compared with Persistence using prediction horizons of 4, 8, 12, and 24 hours. Fiber and Mobile Gateway are included to represent different levels of available backhaul capacity. \Cref{fig:planning_horizon_sensitivity} compares the resulting relay hour savings and fleet action differences across the tested horizons. \Cref{fig:planning_horizon_sensitivity}(a) shows the relay hours saved relative to Persistence, while \Cref{fig:planning_horizon_sensitivity}(b) shows the percentage of control steps where the fleet decision differs by more than 0.1 UAV. \Cref{tab:horizon_sensitivity} reports the backlog reduction, relay hours saved, and the fraction of control steps where the fleet decision differs by more than 0.1 UAV. Positive backlog reduction and relay hour savings indicate an advantage for the Integrated model, while the fraction of different actions measures how often deployment decisions differ.

\begin{table}
\centering
\footnotesize
\renewcommand{\arraystretch}{1.1}
\setlength{\tabcolsep}{3pt}
\begin{tabular}{@{}lrrrr@{}}
\toprule
Backhaul
& Horizon (hour)
& \begin{tabular}[c]{@{}c@{}}Backlog\\reduction (\%)\end{tabular}
& \begin{tabular}[c]{@{}c@{}}Relay hours\\saved (UAV hour)\end{tabular}
& \begin{tabular}[c]{@{}c@{}}Different\\actions (\%)\end{tabular} \\
\midrule
\multirow{4}{*}{Fiber}
& 4  & -0.0041 & 1.09 & 3.8 \\
& 8  & -0.0009 & 2.81 & 12.2 \\
& 12 & -0.0011 & 4.19 & 17.1 \\
& 24 & -0.0006 & 4.69 & 22.1 \\
\midrule
\multirow{4}{*}{Mobile Gateway}
& 4  & 0.0000 & 0.42 & 18.0 \\
& 8  & 0.0000 & 1.13 & 23.0 \\
& 12 & 0.0000 & 1.63 & 32.0 \\
& 24 & 0.0000 & 2.54 & 44.0 \\
\bottomrule
\end{tabular}
\caption{Sensitivity to the prediction horizon for Fiber and Mobile Gateway backhaul.}
\label{tab:horizon_sensitivity}
\end{table}

The backlog results remain effectively unchanged across the tested prediction horizons. Under Fiber, the relative backlog difference remains below 0.005\% for every horizon. The small negative values indicate that the Integrated model produces slightly higher backlog than Persistence, although the differences are small at the scale of the case study. Under Mobile Gateway, the backlog differences remain effectively 0\% (less than $4.1\times10^{-8}\%$) for all tested horizons and are reported as 0\% in \Cref{tab:horizon_sensitivity}. Across the tested prediction horizons, the Integrated model and Persistence produce effectively identical backlog results.

The effect on fleet deployment increases as the prediction horizon becomes longer. In \Cref{fig:planning_horizon_sensitivity}, relay hour savings under Fiber increase from 1.09 UAV hour at a 4 hours horizon to 4.69 UAV hour at a 24 hours horizon, while the fraction of control steps with different fleet decisions increases from 3.8\% to 22.1\%. Under Mobile Gateway, relay hour savings increase from 0.42 to 2.54 UAV hours, and the fraction of different actions increases from 18.0\% to 44.0\%. Relative to the total relay use under Persistence, these savings remain below 0.2\% under Fiber and 0.13\% under Mobile Gateway even at the 24 hours horizon.

These results show that a longer recovery lookahead increasingly changes fleet deployment decisions without a similar improvement in aggregate backlog. Future recovery information contributes primarily to resource efficiency over the tested horizons. The Integrated model can anticipate changes in ground recovery and reduce unnecessary aerial activity, while Persistence achieves nearly identical service performance using only the currently observed recovery state. Furthermore, the planning horizon results provide context for the forecast sensitivity analysis. The limited backlog effect suggests that errors in future recovery information may affect fleet deployment decisions more than the amount of communication service delivered.

\subsubsection{Forecast Error}
\label{sec:exp_sens_forecast}

The forecast error analysis evaluates the sensitivity of the Integrated model to errors in future communication demand and ground recovery information. The current system state is assumed to be observed exactly at each MPC update, and errors are applied only to future values within the prediction horizon. Demand and recovered ground capacity forecasts are perturbed separately using correlated Gaussian errors with a correlation coefficient of 0.70 across future prediction steps within each zone. The resulting multiplicative errors are limited to the range $[-0.75,0.75]$, and demand and ground capacity forecasts are constrained to remain within their physical ranges.

The reported 10\%, 20\%, and 30\% error levels correspond to Gaussian scale parameters of 0.10, 0.20, and 0.30 before clipping. Demand-only and recovery-only errors are evaluated at 20\%, while joint demand and recovery errors are evaluated at 10\%, 20\%, and 30\%. The same standardized error samples are reused across the joint error levels, with only the error scale changed. Each perturbed case uses the same actual demand, ground recovery trajectory, backhaul condition, and initial state as the corresponding perfect forecast case. The perturbations therefore affect only the future information available to the MPC. These synthetic errors are used for sensitivity analysis and are not intended to represent empirically calibrated forecast error distributions.

\begin{table}
\centering
\footnotesize
\renewcommand{\arraystretch}{1.1}
\setlength{\tabcolsep}{5pt}
\begin{tabular}{@{}>{\raggedright\arraybackslash}p{0.25\textwidth}>{\centering\arraybackslash}p{0.10\textwidth}>{\centering\arraybackslash}p{0.16\textwidth}>{\centering\arraybackslash}p{0.16\textwidth}>{\centering\arraybackslash}p{0.22\textwidth}@{}}
\toprule
Forecast input
& Error level (\%)
& Backlog difference (\%)
& Relay use difference (\%)
& Mean fleet decision difference (UAV) \\
\midrule
Demand              & 20 &  0.0148 & 15.8 & 1.59 \\
Recovery            & 20 & -0.0030 & 16.1 & 1.31 \\
Demand and recovery & 10 & -0.0027 &  9.9 & 0.98 \\
Demand and recovery & 20 &  0.0154 & 24.9 & 2.28 \\
Demand and recovery & 30 &  0.0430 & 39.1 & 3.47 \\
\bottomrule
\end{tabular}
\caption{Sensitivity to future demand and recovery forecast errors relative to perfect forecasting. The error level denotes the Gaussian scale parameter before clipping.}
\label{tab:forecast_error}
\end{table}

The results in \Cref{tab:forecast_error} show that forecast errors have little effect on backlog but a much larger effect on UAV use and fleet decisions. Even with 30\% joint demand and recovery errors, backlog increases by only 0.043\%, while relay use increases by 39.1\%. Additionally, the mean fleet decision difference increases with forecast error, reaching 3.47 UAVs at the 30\% joint error level. These results indicate that forecast errors primarily affect UAV resource use rather than communication service under the tested conditions.

\subsection{Computational Performance and Numerical Reliability}
\label{sec:exp_computational}

Computational performance is evaluated across all 5,040 simulation runs, covering seven historical disaster cases, four backhaul scenarios, 30 Monte Carlo samples, and six model cases consisting of the Integrated model and five ablation cases. In \Cref{tab:computational_performance}, values are reported as the mean plus or minus one standard deviation across the simulation runs.
The mean solution time per MPC update is 0.013 plus or minus 0.004 s. The 95th percentile solution time within each run is 0.020 plus or minus 0.012 s, with an overall 95th percentile of 0.027 s. The mean wall clock time for a complete simulation is 11.9 plus or minus 1.2 s, with a median of 11.9 s and a maximum of 18.5 s. These solution times are much shorter than the 1 hour control interval used in the numerical experiments.
\begin{table}[t]
\centering
\footnotesize
\renewcommand{\arraystretch}{1.1}
\begin{tabular}{lr}
\toprule
Metric & Value \\
\midrule
Average solution time per MPC update (s) & $0.013 \pm 0.004$ \\
95th percentile solution time per run (s) & $0.020 \pm 0.012$ \\
Average simulation time per run (s) & $11.9 \pm 1.2$ \\
Failed runs & 0 \\
Average solver fallback rate & $6.97\times10^{-5}$ \\
Maximum constraint violation / tolerance & 0.941 \\
Maximum action adjustment error & $7.96\times10^{-6}$ \\
\bottomrule
\end{tabular}
\caption{Computational performance and numerical reliability.}
\label{tab:computational_performance}
\end{table}

All simulation runs completed successfully, and all accepted solutions satisfied the numerical tolerance. Action projection and energy balance errors were also small. These results show stable numerical performance across the tested disaster, backhaul, and model cases. The solution times are much shorter than the 1 hour control interval used in the numerical experiments, showing that the MPC problems can be solved within the available control time for the tested problem size.

\section{Discussion and Conclusion}
\label{sec:discussion}
This section summarizes the main operational insights from the numerical results, discusses key modeling limitations and directions for future work, and concludes the study.

\subsection{Operational Insights}
\label{sec:op_insights}

The numerical results provide several operational insights for UAV supported communication during disaster recovery.

\textit{1) Dynamic fleet sizing.}
The required level of aerial support changes throughout the recovery period. Cases with slow ground recovery require sustained UAV support, while cases with more advanced recovery require fewer active UAVs. The trajectory results also show that UAV deployment can decrease as ground communication capacity returns. These trajectories show that the number of UAVs needed during disaster recovery is not fixed. A fleet size selected only from the initial disaster condition may lead to unnecessary UAV use later in the recovery period or insufficient support when communication demand remains high. Dynamic fleet sizing allows UAV deployment to follow changes in ground recovery and communication demand.

\textit{2) Backhaul capacity.}
The value of additional aerial resources depends on the available backhaul capacity. Across the tested scenarios, accounting for backhaul constraints reduces the pooled backlog integral by 12.03\% compared with planning that ignores these constraints. The effect of backhaul information is larger under Mobile Gateway and Satellite, whereas the effect of fleet composition is larger under Fiber and Microwave. When backhaul capacity is restrictive, additional UAV relay capacity cannot be fully translated into delivered service. When backhaul capacity is sufficient, communication performance depends more on the fleet's ability to sustain aerial service. These results indicate that UAV fleet requirements and composition should be assessed together with the supporting backhaul infrastructure.

\textit{3) Mobile and Tethered UAVs.}
The mixed fleet performs better than the All Mobile fleet when the same total number of UAVs is available. However, the difference is smaller when the All Mobile fleet includes enough UAVs to provide a similar level of continuous aerial support. This suggests that the main benefit of Tethered UAVs comes from their ability to remain active for long periods without recharging. Mobile UAVs provide flexibility since they can be reassigned across service zones, while Tethered UAVs provide continuous support at fixed locations. A mixed fleet can combine flexible deployment with long duration communication support. From an operational perspective, the choice between Mobile and Tethered UAVs should also consider deployment costs, charging requirements, and site constraints, which are not included in the current model.

\textit{4) Mobile UAV energy.}
The value of Mobile UAV energy modeling depends on communication demand and recharge conditions. Explicit energy modeling has a relatively small effect under the simulated conditions, but its effect increases when recharge takes longer. As communication demand increases, the relative backlog reduction decreases and the absolute backlog reduction increases. Energy modeling becomes more important when repeated flight and recharge cycles change Mobile UAV availability over time. In practice, Mobile UAV fleet size alone does not represent the aerial capacity available over time. Flight and recharge conditions determine how much of the fleet can remain in service during sustained operation.

\textit{5) Recovery and forecast information.}
Current ground recovery information primarily improves resource efficiency in the tested scenarios. Compared with No Recovery, the Integrated model considerably reduces UAV use while producing only a small change in backlog. In contrast, additional information about future ground recovery provides little further benefit at the baseline prediction horizon. The Persistence case produces nearly the same backlog and served ratio, with only a small difference in relay use. Longer prediction horizons lead to more differences in fleet decisions, but these changes provide limited additional savings in UAV use. This suggests that repeatedly updating deployment based on the observed recovery state captures much of the benefit of ground recovery information under the tested conditions. Persistence still uses demand forecasts and receding horizon optimization, so this finding applies specifically to information about future ground recovery.

Forecast errors have a different effect. The tested demand and recovery forecast errors have little effect on backlog but can lead to much larger changes in UAV use and fleet decisions. For example, at the largest tested joint forecast error, backlog changes by less than 0.05\%, while relay use increases by 39.1\%. Similar service performance can therefore require considerably different UAV resource use when forecast accuracy decreases. Forecast quality should be evaluated using both communication performance and UAV resource use.

\subsection{Limitations and Future Work}
\label{sec:limitations}
We make several simplifying assumptions that provide directions for future work. First, we simplify the backhaul model. All service zones use the same backhaul type and capacity within each scenario, and each zone has an independent backhaul connection. Our results show that backhaul capacity can limit the benefit of additional UAVs. Future work could consider different backhaul technologies and capacities across service zones, time varying backhaul availability, and shared backhaul networks. These extensions would allow a more detailed analysis of how backhaul conditions affect the required fleet size and fleet composition.

Second, we simplify the operation of both Mobile and Tethered UAVs. Mobile UAV energy is represented at the fleet level, and deployment decisions represent the average number of active UAVs during each control interval. Tethered UAV locations are fixed, and Mobile UAV routes between service zones are not optimized directly. Moreover, the current framework does not explicitly include deployment and operating costs. Future work could jointly consider Tethered UAV site selection, individual Mobile UAV scheduling and routing, charging resource constraints, travel time, and operational costs.

Finally, additional analysis using observed forecast errors is needed to further evaluate the effect of forecast uncertainty. The current sensitivity analysis uses assumed error levels rather than forecast errors estimated from real disaster data. Future work could use observed demand and recovery forecast errors when such data become available and examine whether the small backlog effects found in the current experiments remain consistent under more realistic forecast conditions. If performance degrades under observed forecast uncertainty, future work could incorporate learning-based demand and recovery prediction before the MPC or introduce a robust MPC formulation to better account for uncertainty.

\subsection{Conclusion}
\label{sec:conclusion}
We developed an MPC framework for dynamic fleet sizing of Mobile and Tethered UAV relays during post-disaster communication recovery. The framework determines the active fleet size and its allocation across service zones while accounting for communication demand, ground network recovery, backhaul capacity, and Mobile UAV energy availability. A conservative piecewise affine approximation of the nonlinear relay model yields a convex quadratic program that supports efficient repeated fleet sizing decisions.

Numerical experiments using seven historical ground recovery trajectories show that UAV support requirements vary substantially across recovery conditions. Persistent communication shortages require sustained aerial deployment, while improving ground network capacity allows the active fleet to decrease over time. These results show that the required UAV fleet is not fixed throughout disaster recovery.

Backhaul availability strongly affects how effectively additional UAV capacity can be translated into communication service. Accounting for backhaul constraints reduces overall backlog by 12.03\% compared with planning that ignores these constraints. When backhaul capacity is high, communication performance depends more on the ability of the UAV fleet to sustain aerial service. Explicit Mobile UAV energy modeling has a smaller effect under the baseline conditions, but its effect increases as recharge time becomes longer.

Current ground recovery information mainly helps reduce unnecessary UAV use, while future recovery information provides little additional backlog reduction over the tested prediction horizons. Forecast errors also affect UAV resource use more than backlog in the evaluated scenarios. Overall, the results show that dynamic UAV fleet sizing should account for both aerial resource availability and the supporting communication infrastructure as recovery conditions change.

\section*{Acknowledgments}
This work was supported by the National Science Foundation IUCRC Phase I: Center for Autonomous Air Mobility and Sensing (CAAMS) under Award No. 2137195.

\section*{CRediT authorship contribution statement}
\textbf{Jimin Choi:} Conceptualization; Methodology; Software; Investigation; Formal analysis; Visualization; Writing -- Original Draft \& Editing.\\
\textbf{Max Z. Li:} Conceptualization; Methodology; Resources; Writing -- Review \& Editing; Supervision; Funding acquisition.

\section*{Declaration of competing interest}

The authors declare that they have no known competing financial interests or personal relationships that could have appeared to influence the work reported in this paper.

\section*{Data availability}

Data will be made available on request.

%% Transportation Research Part C uses an author-year reference style.
\bibliographystyle{elsarticle-num}

\bibliography{cas-refs}

@INPROCEEDINGS{2025bilevel,
  author={Choi, Jimin and Stagg, Grant and Peterson, Cameron K. and Li, Max Z.},
  booktitle={2025 IEEE 64th Conference on Decision and Control (CDC)}, 
  title={Bi-Level Route Optimization and Path Planning with Hazard Exploration}, 
  year={2025},
  volume={},
  number={},
  pages={3403-3410},
  doi={10.1109/CDC57313.2025.11312588}}

@ARTICLE{8910632,
  author={G.C., Deepak and Ladas, Alexandros and Sambo, Yusuf Abdulrahman and Pervaiz, Haris and Politis, Christos and Imran, Muhammad Ali},
  journal={IEEE Wireless Communications}, 
  title={An Overview of Post-Disaster Emergency Communication Systems in the Future Networks}, 
  year={2019},
  volume={26},
  number={6},
  pages={132-139},
  doi={10.1109/MWC.2019.1800467}}

@ARTICLE{9832657,
  author={Matracia, Maurilio and Saeed, Nasir and Kishk, Mustafa A. and Alouini, Mohamed-Slim},
  journal={IEEE Open Journal of the Communications Society}, 
  title={Post-Disaster Communications: Enabling Technologies, Architectures, and Open Challenges}, 
  year={2022},
  volume={3},
  number={},
  pages={1177-1205},
  doi={10.1109/OJCOMS.2022.3192040}}

@ARTICLE{8641424,
  author={Zhao, Nan and Lu, Weidang and Sheng, Min and Chen, Yunfei and Tang, Jie and Yu, F. Richard and Wong, Kai-Kit},
  journal={IEEE Wireless Communications}, 
  title={UAV-Assisted Emergency Networks in Disasters}, 
  year={2019},
  volume={26},
  number={1},
  pages={45-51},
  doi={10.1109/MWC.2018.1800160}}

@INPROCEEDINGS{9013813,
  author={Wu, Di and Sun, Xiang and Ansari, Nirwan},
  booktitle={2019 IEEE Global Communications Conference (GLOBECOM)}, 
  title={A Cooperative Drone Assisted Mobile Access Network for Disaster Emergency Communications}, 
  year={2019},
  volume={},
  number={},
  pages={1-6},
  doi={10.1109/GLOBECOM38437.2019.9013813}}

@Article{app131910723,
AUTHOR = {Gao, Rui and Wang, Xiao},
TITLE = {Rapid Deployment Method for Multi-Scene UAV Base Stations for Disaster Emergency Communications},
JOURNAL = {Applied Sciences},
VOLUME = {13},
YEAR = {2023},
NUMBER = {19},
ARTICLE-NUMBER = {10723},
ISSN = {2076-3417},
DOI = {10.3390/app131910723}
}

@INPROCEEDINGS{8904140,
  author={Mezghani, Farouk and Kortoçi, Pranvera and Mitton, Nathalie and Di Francesco, Mario},
  booktitle={2019 IEEE 30th Annual International Symposium on Personal, Indoor and Mobile Radio Communications (PIMRC)}, 
  title={A Multi-tier Communication Scheme for Drone-assisted Disaster Recovery Scenarios}, 
  year={2019},
  volume={},
  number={},
  pages={1-7},
  doi={10.1109/PIMRC.2019.8904140}}

@article{DEFRAEYE20164,
title = {Staffing and scheduling under nonstationary demand for service: A literature review},
journal = {Omega},
volume = {58},
pages = {4-25},
year = {2016},
issn = {0305-0483},
doi = {10.1016/j.omega.2015.04.002},
author = {Mieke Defraeye and Inneke {Van Nieuwenhuyse}},
}

@article{SEIDL2016246,
title = {Optimal control of a terror queue},
journal = {European Journal of Operational Research},
volume = {248},
number = {1},
pages = {246-256},
year = {2016},
issn = {0377-2217},
doi = {10.1016/j.ejor.2015.07.010},
author = {Andrea Seidl and Edward H. Kaplan and Jonathan P. Caulkins and Stefan Wrzaczek and Gustav Feichtinger},
}

@Article{su10093344,
AUTHOR = {Troudi, Asma and Addouche, Sid-Ali and Dellagi, Sofiene and Mhamedi, Abderrahman El},
TITLE = {Sizing of the Drone Delivery Fleet Considering Energy Autonomy},
JOURNAL = {Sustainability},
VOLUME = {10},
YEAR = {2018},
NUMBER = {9},
ARTICLE-NUMBER = {3344},
ISSN = {2071-1050},
DOI = {10.3390/su10093344}
}

@article{HADAS2024100127,
title = {Modeling optimal drone fleet size considering stochastic demand},
journal = {EURO Journal on Transportation and Logistics},
volume = {13},
pages = {100127},
year = {2024},
issn = {2192-4376},
doi = {10.1016/j.ejtl.2024.100127},
author = {Yuval Hadas and Miguel A. Figliozzi},
}

@article{PEI2021102482,
title = {Managing price and fleet size for courier service with shared drones},
journal = {Omega},
volume = {104},
pages = {102482},
year = {2021},
issn = {0305-0483},
doi = {10.1016/j.omega.2021.102482},
author = {Zhi Pei and Xu Dai and Yilun Yuan and Rui Du and Changchun Liu},
}

@techreport{fcc2018AtlanticHurricanes,
  author      = {{Federal Communications Commission, Public Safety and Homeland Security Bureau}},
  title       = {2017 Atlantic Hurricane Season Impact on Communications: Report and Recommendations},
  institution = {Federal Communications Commission},
  type        = {Report},
  number      = {Public Safety Docket No. 17-344},
  year        = {2018},
  month       = aug,
  url         = {https://docs.fcc.gov/public/attachments/DOC-353805A1.pdf}
}

@techreport{fcc2019HurricaneMichael,
  author      = {{Federal Communications Commission, Public Safety and Homeland Security Bureau}},
  title       = {October 2018 Hurricane Michael's Impact on Communications: Preparation, Effect, and Recovery: Report and Recommendations},
  institution = {Federal Communications Commission},
  type        = {Report},
  number      = {Public Safety Docket No. 18-339},
  year        = {2019},
  month       = may,
  url         = {https://docs.fcc.gov/public/attachments/DOC-357387A1.pdf}
}

@techreport{fcc2023HawaiiWildfires,
  author      = {{Federal Communications Commission}},
  title       = {Communications Status Report for Areas Impacted by Hawaii Wildfires, August 23, 2023},
  institution = {Federal Communications Commission},
  type        = {Communications Status Report},
  year        = {2023},
  month       = aug,
  note        = {23 August 2023},
  url         = {https://docs.fcc.gov/public/attachments/DOC-396290A1.pdf}
}

@techreport{fcc2024HurricaneHelene,
  author      = {{Federal Communications Commission}},
  title       = {Communications Status Report for Areas Impacted by Hurricane Helene, October 1, 2024},
  institution = {Federal Communications Commission},
  type        = {Communications Status Report},
  year        = {2024},
  month       = oct,
  note        = {1 October 2024},
  url         = {https://docs.fcc.gov/public/attachments/DOC-406055A1.pdf}
}

@techreport{itu2008g9841,
  author      = {{International Telecommunication Union}},
  title       = {Gigabit-Capable Passive Optical Networks ({GPON}): General Characteristics},
  institution = {International Telecommunication Union},
  type        = {{ITU-T} Recommendation},
  number      = {G.984.1},
  year        = {2008},
  month       = mar,
  url         = {https://www.itu.int/rec/T-REC-G.984.1-200803-I}
}

@misc{cambiumPTP820S,
  author       = {{Cambium Networks}},
  title        = {{PTP 820S} Specifications},
  howpublished = {Official product specification},
  note         = {Accessed 17 September 2026},
  url          = {https://www.cambiumnetworks.com/resource/ptp-820s-specs/}
}

@manual{peplinkBR2Pro,
  author       = {{Peplink}},
  title        = {{BR2 Pro} Technical Specifications},
  year         = {2026},
  note         = {Accessed 17 September 2026},
  url          = {https://www.peplink.com/compare/tech-specs/br2-pro.pdf}
}

@misc{starlinkSpecifications,
  author       = {{Starlink}},
  title        = {Starlink Specifications},
  howpublished = {Official service performance specifications},
  note         = {Accessed 17 September 2026},
  url          = {https://starlink.com/legal/documents/DOC-1470-99699-90}
}

@misc{pdwC100,
  author       = {{Performance Drone Works (PDW)}},
  title        = {{C100 Defense}: Multi-Mission Tactical {UAS}},
  howpublished = {Official product specification},
  note         = {Accessed 17 September 2026},
  url          = {https://www.pdw.ai/products/c100-defense}
}

@misc{elistairKhronos,
  author       = {{Elistair}},
  title        = {{Khronos}: Advanced Tethered DroneBox},
  howpublished = {Official product specification},
  note         = {Accessed 17 September 2026},
  url          = {https://elistair.com/solutions/tethered-dronebox-khronos/}
}

@misc{firstnetDeployables,
  author       = {{First Responder Network Authority}},
  title        = {Deployables},
  howpublished = {FirstNet Authority},
  url          = {https://firstnet.gov/network/TT/deployables},
  note         = {Accessed: 2026-09-18}
}

@article{liXu2018Backhaul,
  author  = {Li, Peiming and Xu, Jie},
  title   = {Placement Optimization for UAV-Enabled Wireless Networks with Multi-Hop Backhauls},
  journal = {Journal of Communications and Information Networks},
  year    = {2018},
  volume  = {3},
  pages   = {64--73},
  doi     = {10.1007/s41650-018-0040-3},
  url     = {https://doi.org/10.1007/s41650-018-0040-3}
}

@misc{att2024Helene,
  author       = {{AT\&T}},
  title        = {{AT\&T} Supports Communities Affected by Hurricane Helene},
  year         = {2024},
  month        = oct,
  howpublished = {\url{https://about.att.com/pages/disaster-recovery/2024/storm-helene}},
  note         = {Accessed October 1, 2026}
}

@article{zeng2016uav,
  author  = {Zeng, Yong and Zhang, Rui and Lim, Teng Joon},
  title   = {Wireless Communications with Unmanned Aerial Vehicles: Opportunities and Challenges},
  journal = {IEEE Communications Magazine},
  year    = {2016},
  volume  = {54},
  number  = {5},
  pages   = {36--42},
  month   = may,
  doi     = {10.1109/MCOM.2016.7470933}
}

@article{naqvi2018drone,
  author  = {Naqvi, Syed Ahsan Raza and Hassan, Syed Ali and Pervaiz, Haris and Ni, Qiang},
  title   = {Drone-Aided Communication as a Key Enabler for 5G and Resilient Public Safety Networks},
  journal = {IEEE Communications Magazine},
  year    = {2018},
  volume  = {56},
  number  = {1},
  pages   = {36--42},
  month   = jan,
  doi     = {10.1109/MCOM.2017.1700451}
}

@inproceedings{hayajneh2016drone,
  author    = {Hayajneh, Ali Mohammad and Zaidi, Syed Ali Raza and McLernon, Des C. and Ghogho, Mounir},
  title     = {Drone Empowered Small Cellular Disaster Recovery Networks for Resilient Smart Cities},
  booktitle = {2016 IEEE International Conference on Sensing, Communication and Networking (SECON Workshops)},
  year      = {2016},
  pages     = {1--6},
  doi       = {10.1109/SECONW.2016.7746806}
}

@techreport{firstnet2023Maui,
  author      = {{First Responder Network Authority}},
  title       = {FirstNet Authority FY2023 Annual Report},
  institution = {First Responder Network Authority},
  year        = {2023},
  note        = {\url{https://firstnet.gov/sites/default/files/FirstNetAuthority_AnnualReport_FY2023.pdf}}
}

@misc{att2024Beryl,
  author       = {{AT\&T}},
  title        = {{AT\&T} Networks Prepare for Hurricane Beryl},
  year         = {2024},
  month        = jul,
  howpublished = {\url{https://about.att.com/pages/disaster_relief/2024/storm-beryl.html}},
  note         = {Accessed October 1, 2026}
}

@article{uavEnergy2024,
  author  = {Panahi, Farzad H. and Panahi, Fereidoun H.},
  title   = {Reliable and Energy-Efficient UAV Communications: A Cost-Aware Perspective},
  journal = {IEEE Transactions on Mobile Computing},
  volume  = {23},
  number  = {5},
  pages   = {4038--4049},
  year    = {2024},
  month   = may,
  doi     = {10.1109/TMC.2023.3284531}
}

@article{tetheredUAVReview2025,
  author  = {Fattori, Francesco and Cocuzza, Silvio},
  title   = {Tethered Drones: A Comprehensive Review of Technologies, Challenges, and Applications},
  journal = {Drones},
  volume  = {9},
  number  = {6},
  pages   = {425},
  year    = {2025},
  month   = jun,
  doi     = {10.3390/drones9060425}
}

@article{uavBackhaul2020,
  author  = {Kumar, Sidharth and Suman, Suraj and De, Swades},
  title   = {Dynamic Resource Allocation in UAV-Enabled mmWave Communication Networks},
  journal = {IEEE Internet of Things Journal},
  volume  = {8},
  number  = {12},
  pages   = {9920--9933},
  year    = {2021},
  month   = jun,
  doi     = {10.1109/JIOT.2020.3027476}
}

@inproceedings{choi2026optimal,
  author    = {Choi, Jimin and Li, Max Z.},
  title     = {Optimal Control-Based Adaptive Fleet Sizing of UAV Communication Relays for Post-Disaster Network Connectivity Recovery},
  booktitle = {AIAA AVIATION 2026 Forum},
  year      = {2026},
  month     = jun,
  doi       = {10.2514/6.2026-4335}
}

@article{ghelichi2022drone,
  author  = {Ghelichi, Zabih and Gentili, Monica and Mirchandani, Pitu B.},
  title   = {Drone Logistics for Uncertain Demand of Disaster-Impacted Populations},
  journal = {Transportation Research Part C: Emerging Technologies},
  volume  = {141},
  pages   = {103735},
  year    = {2022},
  doi     = {10.1016/j.trc.2022.103735}
}

@article{vanSteenbergen2023relief,
  author  = {van Steenbergen, Robert and Mes, Martijn and van Heeswijk, Wouter},
  title   = {Reinforcement Learning for Humanitarian Relief Distribution with Trucks and UAVs under Travel Time Uncertainty},
  journal = {Transportation Research Part C: Emerging Technologies},
  volume  = {157},
  pages   = {104401},
  year    = {2023},
  doi     = {10.1016/j.trc.2023.104401}
}

@article{choi2026airground,
  author  = {Choi, Jimin and Li, Max Z.},
  title   = {Autonomous Air--Ground Vehicle Operations Optimization in Hazardous Environments: A Multi-Armed Bandit Approach},
  journal = {Journal of Aerospace Information Systems},
  volume  = {23},
  pages   = {694--709},
  year    = {2026},
  doi     = {10.2514/1.I011749}
}

@article{liu2019stochasticFleet,
  author  = {Liu, Ming and Liu, Xin and Zhu, Maoran and Zheng, Feifeng},
  title   = {Stochastic Drone Fleet Deployment and Planning Problem Considering Multiple-Type Delivery Service},
  journal = {Sustainability},
  volume  = {11},
  number  = {14},
  pages   = {3871},
  year    = {2019},
  month   = jul,
  doi     = {10.3390/su11143871}
}

@inproceedings{hartuv2018spareDrones,
  author    = {Hartuv, Erez and Agmon, Noa and Kraus, Sarit},
  title     = {Scheduling Spare Drones for Persistent Task Performance under Energy Constraints},
  booktitle = {Proceedings of the 17th International Conference on Autonomous Agents and Multiagent Systems},
  pages     = {532--540},
  year      = {2018},
  doi       ={10.5555/3237383.3237463}
}

@article{gao2013dynamicResource,
  author  = {Gao, Xuefeng and Lu, Yingdong and Sharma, Mayank and Squillante, Mark S. and Bosman, J. W.},
  title   = {Stochastic Optimal Control for a General Class of Dynamic Resource Allocation Problems},
  journal = {ACM SIGMETRICS Performance Evaluation Review},
  volume  = {41},
  number  = {2},
  pages   = {3--14},
  year    = {2013},
  doi     = {10.1145/2518025.2518027}
}

@article{powell1987dynamicAllocation,
  author  = {Powell, Warren B.},
  title   = {An Operational Planning Model for the Dynamic Vehicle Allocation Problem with Uncertain Demands},
  journal = {Transportation Research Part B: Methodological},
  volume  = {21},
  number  = {3},
  pages   = {217--232},
  year    = {1987},
  month   = jun,
  doi     = {10.1016/0191-2615(87)90005-1}
}

@article{hajibabai2016snowPlow,
  author  = {Hajibabai, Leila and Ouyang, Yanfeng},
  title   = {Dynamic Snow Plow Fleet Management under Uncertain Demand and Service Disruption},
  journal = {IEEE Transactions on Intelligent Transportation Systems},
  volume  = {17},
  number  = {9},
  pages   = {2574--2582},
  year    = {2016},
  month   = sep,
  doi     = {10.1109/TITS.2016.2520918}
}

@article{milenkovic2015mpcFleet,
  author  = {Milenkovi{\'c}, Milo{\v{s}} S. and Bojovi{\'c}, Neboj{\v{s}}a J. and {\v{S}}vadlenka, Libor and Melichar, Vlastimil},
  title   = {A Stochastic Model Predictive Control to Heterogeneous Rail Freight Car Fleet Sizing Problem},
  journal = {Transportation Research Part E: Logistics and Transportation Review},
  volume  = {82},
  pages   = {162--198},
  year    = {2015},
  month   = oct,
  doi     = {10.1016/j.tre.2015.07.009}
}

@article{deng2026heterogeneousUAV,
  author  = {Deng, Jikang and Zhou, Hui and Alouini, Mohamed-Slim},
  title   = {Two-Timescale Optimization Framework for {IAB}-Enabled Heterogeneous {UAV} Networks},
  journal = {IEEE Internet of Things Journal},
  volume  = {13},
  number  = {2},
  pages   = {3536--3551},
  year    = {2026},
  doi     = {10.1109/JIOT.2025.3634466}
}

@article{auadPerez2022,
  author  = {Auad-Perez, Ramon and Van Hentenryck, Pascal},
  title   = {Ridesharing and Fleet Sizing for On-Demand Multimodal Transit Systems},
  journal = {Transportation Research Part C: Emerging Technologies},
  volume  = {138},
  pages   = {103594},
  year    = {2022},
  doi     = {10.1016/j.trc.2022.103594}
}

@article{lu2016bike,
  author  = {Lu, Chung-Cheng},
  title   = {Robust Multi-period Fleet Allocation Models for Bike-Sharing Systems},
  journal = {Networks and Spatial Economics},
  volume  = {16},
  number  = {1},
  pages   = {61--82},
  year    = {2016},
  doi     = {10.1007/s11067-013-9203-9}
}

@article{cimellaro2010resilience,
  author  = {Cimellaro, Gian Paolo and Reinhorn, Andrei M. and Bruneau, Michel},
  title   = {Framework for Analytical Quantification of Disaster Resilience},
  journal = {Engineering Structures},
  volume  = {32},
  number  = {11},
  pages   = {3639--3649},
  year    = {2010},
  doi     = {10.1016/j.engstruct.2010.08.008}
}

@article{zorn2015restoration,
  author  = {Zorn, Conrad R. and Shamseldin, Asaad Y.},
  title   = {Post-Disaster Infrastructure Restoration: A Comparison of Events for Future Planning},
  journal = {International Journal of Disaster Risk Reduction},
  volume  = {13},
  pages   = {158--166},
  year    = {2015},
  doi     = {10.1016/j.ijdrr.2015.04.004}
}

@article{arribas2023resupply,
  author  = {Arribas, Edgar and Cholvi, Vicent and Mancuso, Vincenzo},
  title   = {Optimizing UAV Resupply Scheduling for Heterogeneous and Persistent Aerial Service},
  journal = {IEEE Transactions on Robotics},
  volume  = {39},
  number  = {4},
  pages   = {2639--2653},
  year    = {2023},
  doi     = {10.1109/TRO.2023.3263077}
}

@article{song2014persistent,
  author  = {Song, Byung Duk and Kim, Jonghoe and Kim, Jeongwoon and Park, Hyorin and Morrison, James R. and Shim, David Hyunchul},
  title   = {Persistent UAV Service: An Improved Scheduling Formulation and Prototypes of System Components},
  journal = {Journal of Intelligent \& Robotic Systems},
  volume  = {74},
  number  = {1--2},
  pages   = {221--232},
  year    = {2014},
  doi     = {10.1007/s10846-013-9970-z}
}

\end{document}